\documentclass[superscriptaddress,citeautoscript,floatfix,longbibliography,nofootinbib]{revtex4-2}
\usepackage{times,amsmath,amsfonts,amsthm,amssymb,amsbsy,bm,braket,graphicx}
\usepackage{hyperref}
\hypersetup{colorlinks=true,urlcolor=blue,linkcolor=blue,citecolor=red}
\usepackage[utf8]{inputenc}
\usepackage[T1]{fontenc}
\usepackage[capitalize]{cleveref}
\usepackage{microtype}
\usepackage{mhchem}
\usepackage[misc]{ifsym}
\let\oldcite\cite\renewcommand{\cite}{~\oldcite}
\newcommand{\ii}{\mathrm{i}}
\newcommand{\ee}{\mathrm{e}}
\newcommand{\dd}{\mathrm{d}}
\newcommand{\up}{\uparrow}
\newcommand{\down}{\downarrow}
\DeclareMathOperator{\sgn}{sgn}
\DeclareMathOperator{\pf}{pf}

\makeatletter
\def\@fnsymbol#1{{\Letter}}
\makeatother

\begin{document}
\title{
Majorana modes in striped two-dimensional inhomogeneous topological superconductors
}

\author{Pasquale Marra}
\email[corresponding author: ]{pmarra@ms.u-tokyo.ac.jp}
\affiliation{
Graduate School of Mathematical Sciences,
The University of Tokyo, 3-8-1 Komaba, Meguro, Tokyo, 153-8914, Japan
}
\affiliation{
Department of Physics \& Research and Education Center for Natural Sciences, 
Keio University, 4-1-1 Hiyoshi, Yokohama, Kanagawa, 223-8521, Japan
}
\author{Daisuke Inotani} 
\affiliation{
Department of Physics \& Research and Education Center for Natural Sciences, 
Keio University, 4-1-1 Hiyoshi, Yokohama, Kanagawa, 223-8521, Japan
}
\author{Takeshi Mizushima}
\affiliation{
Department of Materials Engineering Science, Osaka University, Toyonaka, Osaka 560-8531, Japan
}
\author{Muneto Nitta} 
\affiliation{
Department of Physics \& Research and Education Center for Natural Sciences, 
Keio University, 4-1-1 Hiyoshi, Yokohama, Kanagawa, 223-8521, Japan
}
\affiliation{
International Institute for Sustainability with Knotted Chiral Meta Matter (SKCM$^2$), Hiroshima University, 1-3-2 Kagamiyama, Higashi-Hiroshima, Hiroshima, 739-8511, Japan
}
\begin{abstract}
Majorana zero modes have gained significant interest due to their potential applications in topological quantum computing and in the realization of exotic quantum phases. These zero-energy quasiparticle excitations localize at the vortex cores of two-dimensional topological superconductors or at the ends of one-dimensional topological superconductors. Here we describe an alternative platform: a two-dimensional topological superconductor with inhomogeneous superconductivity, where Majorana modes localize at the ends of topologically nontrivial one-dimensional stripes induced by the spatial variations of the order parameter phase. In certain regimes, these Majorana modes hybridize into a single highly nonlocal state delocalized over spatially separated points, with exactly zero energy at finite system sizes and with emergent quantum-mechanical supersymmetry. We then present detailed descriptions of braiding and fusion protocols and showcase the versatility of our proposal by suggesting possible setups that can potentially lead to the realization of Yang-Lee anyons and the Sachdev-Ye-Kitaev model.
\end{abstract}
\date\today
\maketitle

\section*{Introduction}

Majorana modes (MMs) localized at the vortex cores of two-dimensional (2D) or at the ends of one-dimensional (1D) topological superconductors (TSs)\cite{read_paired_2000,kitaev_unpaired_2001,fu_superconducting_2008,sato_non-abelian_2009,sato_topological_2009,lutchyn_majorana_2010,oreg_helical_2010} are potential building blocks for topological quantum computing\cite{ivanov_non-abelian_2001,kitaev_fault-tolerant_2003,stern_geometric_2004,nayak_non-abelian_2008} and other exotic quantum systems that effectively simulate high-energy theories such as supersymmetry (SUSY)\cite{rahmani_phase_2015,rahmani_emergent_2015,hsieh_all-majorana_2016,huang_supersymmetry_2017,sannomiya_supersymmetry_2019,marra_1d-majorana_2022,marra_dispersive_2022,miura_interacting_2024} and synthetic horizons in the Sachdev-Ye-Kitaev (SYK) model\cite{sachdev_gapless_1993,kitaev_a-simple_2015,chew_approximating_2017} (see also Refs.~\onlinecite{alicea_new-directions_2012,leijnse_introduction_2012,tanaka_symmetry_2012,stanescu_majorana_2013,beenakker_search_2013,elliott_colloquium:_2015,das-sarma_majorana_2015,sato_majorana_2016,sato_topological_2017,aguado_majorana_2017,laubscher_majorana_2021,marra_majorana_2022,masaki_non-abelian_2024,tanaka_theory_2024}).
Specifically, 2D TSs\cite{read_paired_2000} realized with topological insulator (TI) or quantum spin Hall insulator/superconductor heterostructures\cite{fu_superconducting_2008}, transition metal dichalcogenides\cite{yuan_possible_2014,hsu_topological_2017}, 
iron pnictides\cite{yin_observation_2015,zhang_observation_2018,wang_evidence_2018} 
or other oxypnictide superconductors\cite{huang_dual_2022}, or magnet-superconductor hybrid systems\cite{mascot_topological_2021,crawford_majorana_2022,escribano_semiconductor-ferromagnet-superconductor_2022,crawford_majorana_2022,wong_higher_2023}
may provide a flexible platform to exploit the nonabelian exchange statistics of MMs\cite{ivanov_non-abelian_2001}, due to the ability to manipulate vortex cores in a 2D space, and detect them through scanning tunneling microscopy (STM).
Conversely, 1D TSs\cite{kitaev_unpaired_2001}, e.g., proximitized quantum wires\cite{lutchyn_majorana_2010,oreg_helical_2010,sau_non-abelian_2010,akhmerov_quantized_2011,stanescu_majorana_2011} 
arranged in 2D networks, offer a simplified but less flexible setup to perform braiding\cite{alicea_non-abelian_2011,clarke_majorana_2011,halperin_adiabatic_2012}.

To combine the flexibility of braiding in 2D with the conceptual simplicity of 1D platforms and overcome their limitations, here we introduce topologically nontrivial stripes (TNSs) induced by
inhomogeneous superconducting states\cite{fulde_superconductivity_1964,larkin_nonuniform_1964,casalbuoni_inhomogeneous_2004} where the gauge-invariant phase rotates in a regular pattern. 
The rotating phase effectively generates quasi-1D structures within the 2D system, where the topological invariant assumes alternatively trivial and nontrivial values as a function of the phase.
This results into a \emph{striped} 2D TS with emergent TNSs equivalent to 1D TSs, localized at 1D lines where the order parameter phase is homogeneous and the topological invariant is nontrivial, so that a quasi-1D topological superconducting state emerges.
Here, pointlike (0D) MMs localize at the ends of the TNSs, whose distance and direction can be manipulated by varying the in-plane field magnitude and direction.
These highly nonlocal MMs offer multiple and flexible ways to implement braiding due to the possibility of moving and rotating the stripes in a 2D space. 

Specifically, we consider a TI film in a magnetic field\cite{lu_massive_2010,yu_quantized_2010,hu_topological_2019} where the surface states are gapped out by proximity with a conventional superconductor, allowing the realization of second-order TSs\cite{schindler_higher-order_2018} with Majorana hinge modes.
We show that these (1D) Majorana hinge modes transmute into pointlike (0D) Majorana corner modes localized at the end of TNSs induced by an inhomogeneous superconducting order.
Hence, we describe braiding and fusion protocols implemented by joining, splitting, and moving stripes via external gates or magnetic force microscopy, and rotating them by rotating the magnetic field.
Furthermore, we show how configurations with several stripes induce a regularly-spaced array of MMs, realizing emergent quantum mechanical SUSY, zero-energy multi-locational MMs\cite{nagae_multilocational_2024} delocalized on multiple spatially separated points, Yang-Lee anyons\cite{sanno_engineering_2022} with non-unitary and nonabelian statistics, and the SYK model\cite{chew_approximating_2017} reproducing the maximally-chaotic dynamics of black holes.

\begin{figure*}[t]
\centering
\includegraphics[width=\textwidth]{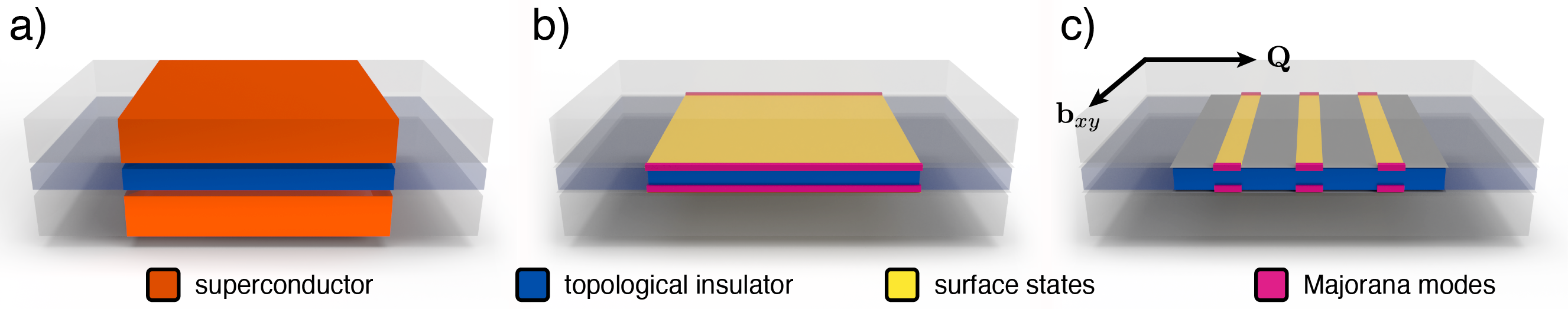} 
\caption{
Topological insulator film with top and bottom surfaces proximitized by conventional superconductors.
(a) Topological insulator film sandwiched between two conventional superconductors.
(b) Majorana hinge modes in the nontrivial phase at finite field and uniform pairing (we removed the superconductors for clarity).
(c) Topologically nontrivial stripes and Majorana corner modes in the nonuniform superconducting phase.
The LO order parameter is modulated as $\Delta(\mathbf{r})\propto\ee^{\ii \mathbf{Q}\cdot\mathbf{r}}$.
The Cooper pairs momentum $\mathbf{Q}$ is perpendicular to the in-plane magnetic field $\mathbf{b}_{xy}$.
}
\label{fig:system}
\end{figure*}

\section*{Results}

\subsection*{Creation of topologically nontrivial stripes}

The boundary mean-field Hamiltonian describing the proximitized surface states at the top and bottom of a TI in a magnetic field as sketched in \cref{fig:system}(a) is\cite{lu_massive_2010,yu_quantized_2010,zhang_topological_2013,hu_topological_2019} 
\begin{align}
\mathcal{H} =& \!\int\!\dd\mathbf{r}\, 
{\psi}^\dag(\mathbf{r}) H{\psi}(\mathbf{r})
-
\sum_{\tau}
\left(
\Delta_{\tau}(\mathbf{r})
{\psi}_{\tau\down}(\mathbf{r})
{\psi}_{\tau\up}(\mathbf{r})
+\text{h.c.}
\right)
\nonumber\\&+
\frac1U\int\!\dd\mathbf{r}\,\sum_{\tau} 
|\Delta_{\tau}(\mathbf{r})|^2
,
\end{align}
where 
${\psi}(\mathbf{r})=\left[{\psi}_{1,\up}(\mathbf{r}),{\psi}_{1,\down}(\mathbf{r}),{\psi}_{2,\up}(\mathbf{r}),{\psi}_{2,\down}(\mathbf{r})\right]^\intercal$ 
are the electron field operators with $\up\down$ indexes for spin, $\tau=1,2$ for pseudospin (i.e., top and bottom surfaces) and 
\begin{equation}\label{eq:H}
H= 
( m' \mathbf{p}^2 + m)\tau_x - v\mathbf{p}\times\bm{\sigma}\tau_z
-\mu
+ \mathbf{b}\cdot\bm{\sigma},
\end{equation}
where
$m$ and $m'$ are the tunneling parameters depending on the layer thickness\cite{lu_massive_2010,liu_oscillatory_2010},
$v$ the Dirac cone velocity,
$\mathbf{p}$ the momentum on $xy$ plane, 
$\mathbf{b}$ the Zeeman field,
$\mu$ the chemical potential,
and $\sigma_i$, $\tau_i$ the Pauli matrices in spin and pseudospin space.
Here,
$\Delta_{\tau}(\mathbf{r})$
is the mean-field order parameter of the surface states, i.e., at the interface between the superconductor and the TI, and
which can be calculated self-consistently\cite{black-schaffer_self-consistent_2008,lababidi_microscopic_2011,hu_topological_2019}
as
$\Delta_{\tau}(\mathbf{r})=
-U
\bra{\text{0}}
{\psi}_{\tau\down}(\mathbf{r})
{\psi}_{\tau\up}(\mathbf{r})
\ket{\text{0}}$ 
,
where $U$ the superconducting pairing strength describing the on-site electron-electron attractive interaction within the surface states.
In the normal regime with unbroken time-reversal symmetry (i.e., $\Delta=b=0$), the Hamiltonian is unitarily equivalent\cite{lu_massive_2010,liu_oscillatory_2010} to the Bernevig, Hughes, and Zhang model for a 2D quantum spin Hall TI\cite{bernevig_quantum_2006a} with energy dispersion having two spin-degenerate branches with gap $2|m|$ at zero momentum and a topologically nontrivial phase for $m m'<0$.
The TI surface states exhibit a gap $2(b-|m|)$ at zero momentum for $b\neq|m|$ and remain gapped at finite momenta for finite out-of-plane fields $b_z>0$.

Superconductivity emerges when the Cooper instability opens a gap at the Fermi level lying within the conduction band, assuming $|\mu|>b-|m|$.
The symmetry between the top and bottom surfaces mandates $|\Delta_1(\mathbf{r})|=|\Delta_2(\mathbf{r})|$. 
Hence, by choosing a gauge where the phases of the order parameter on the top and bottom surfaces are opposite, one can write $\Delta_1(\mathbf{r})=\Delta_2(\mathbf{r})^*$ up to a gauge transformation.
We can thus write $\Delta_1(\mathbf{r})=\Delta(\mathbf{r})=|\Delta(\mathbf{r})|\ee^{\ii\phi(\mathbf{r})}$ and $\Delta_2(\mathbf{r})=\Delta(\mathbf{r})^*=|\Delta(\mathbf{r})|\ee^{-\ii\phi(\mathbf{r})}$, where $2\phi(\mathbf{r})$ is the gauge-invariant phase difference between top and bottom surfaces.
The mean-field Bogoliubov-de~Gennes Hamiltonian is
$\mathcal{H}_\mathrm{BdG}=
\frac12\int \dd\mathbf{r}\,\Psi^\dag(\mathbf{r})\cdot H_\mathrm{BdG}\cdot\Psi(\mathbf{r})$
with
\begin{gather}
H_\mathrm{BdG}=
\left[
( m' \mathbf{p}^2 + m)\tau_x
-v\mathbf{p}\times\bm{\sigma}\tau_z
-\mu
\right]
\upsilon_z 
+
\mathbf{b}\cdot\bm{\sigma}
\nonumber\\+
|\Delta(\mathbf{r})|\left[
\cos(\phi(\mathbf{r}))\,\upsilon_x+
\sin(\phi(\mathbf{r}))\,\tau_z\upsilon_y
\right]
,
\label{eq:HBdG}
\end{gather}
where $\Psi(\mathbf{r})=\left[\psi(\mathbf{r}),\sigma_y\psi^\dag(\mathbf{r})\sigma_y \right]^\intercal$, and $\upsilon_i$ the Pauli matrices in particle-hole space.

Let us first assume uniform superconducting pairing $\Delta(\mathbf{r})=\Delta\ee^{\ii\phi}$ with $\Delta>0$.
Topologically nontrivial phases with particle-hole symmetry and broken time-reversal symmetry (class D) in 2D are labeled by the Chern number of the 
quasiparticle excitation gap $c\in\mathbb{Z}$.
The gap closes when $|m^2+\mu^2+\Delta^2-b^2| =2 |m| \sqrt{\mu^2+\Delta ^2 \sin^2{\phi}}$, and remains open at finite momenta for $b_z\neq0$ and $\phi\neq0$.
For $\phi=\pi/2$, the quasiparticle excitation gap $2\min(| |m|-| b\pm\sqrt{\mu^2+\Delta^2}| | )$ closes at zero momentum with a quantum phase transition each time that any of the quantities $b\pm m\pm\sqrt{\mu^2+\Delta^2}$ change sign.
This condition divides the parameter space into topologically distinct phases separated by the closing of the quasiparticle excitation gap,
where we calculate the Chern number numerically\cite{fukui_chern_2005}:
We thus found a trivial phase at weak fields, where $|m|>b+\sqrt{\mu^2+\Delta^2}$ or $\sqrt{\mu^2+\Delta^2}>b+|m|$, a nontrivial phase with $|c|=2$ at strong field $b>|m|+\sqrt{\mu^2+\Delta^2}$,
and a nontrivial intermediate phase with $|c|=1$ where no energy scale dominates, i.e., when 
$|m|+\sqrt{\mu^2+\Delta^2}>b>| |m|-\sqrt{\mu^2+\Delta^2} |$, or equivalently
$b+|m|>\sqrt{\mu^2+\Delta^2}>| b-|m| |$, or 
$b+\sqrt{\mu^2+\Delta^2}>|m|>| b-\sqrt{\mu^2+\Delta^2} |$.
The nontrivial phases persist for $\phi\neq\pi/2$ as long as the quasiparticle excitation gap remains open (see also Supplementary Note 2).
The parity of the topological invariant\cite{tewari_topological_2012a,budich_equivalent_2013} $\nu=c\mod 2$ is given by $(-1)^{\nu}=\prod_{\mathbf{k}}\sgn\left(\pf\left(H_\mathrm{BdG}(\mathbf{k})\sigma_y\upsilon_y\right)\right)$ where $H_\mathrm{BdG}(\mathbf{k})$ is the Hamiltonian density as a function of the momentum eigenvalues $k$ with the product spanning over the time-reversal symmetry points of the Brillouin zone, giving
\begin{equation}
(-1)^\nu=\sgn
\left(
\left\vert m^2+\mu^2+\Delta^2-b^2\right\vert
-2 |m| \sqrt{\mu^2+\Delta^2 \sin^2{\phi}}
\right).
\label{eq:Z2invariant}
\end{equation}
Since the effective boundary Hamiltonian in \cref{eq:H} describes the surface states of a 3D TI, these nontrivial gapped phases are 3D second-order TS with Majorana hinge modes, i.e., gapless modes on the hinges\cite{schindler_higher-order_2018}, as in \cref{fig:system}(b).

For zero magnetic fields or fields parallel to the $z$-axis, the $\mathrm{SO}(2)$ rotational symmetry in the $xy$ plane is unbroken:
this allows the creation of Cooper pairs with zero momentum $Q=0$ formed by electrons with opposite spin and opposite momenta.
However, 
in the presence of a finite spin-orbit coupling term $\propto\mathbf{p}\times\bm{\sigma}$, a finite
in-plane magnetic field $\mathbf{b}\cdot\bm{\sigma}$ (Zeeman term) shift electrons 
with opposite spin in opposite directions $\mathbf{k}\to\mathbf{k}\pm\mathbf{Q}/2$, with the momentum $\mathbf{Q}$ perpendicular to the in-plane field and $Q\approx 2b_{xy}/v$ at large fields.
In the Pauli limit, neglecting the orbital pair-breaking mechanism, this allows the creation of Cooper pairs with finite momentum $\mathbf{Q}$, formed by electrons with opposite spin and momentum eigenvalues $\mathbf{k}$ and $-\mathbf{k}+\mathbf{Q}$, described by a nonuniform order parameter $\Delta(\mathbf{r})$ that depends periodically in space with a wavelength $\lambda=2\pi/Q$.
The simplest spatial dependence compatible with the symmetries of the system considered here\cite{hu_topological_2019} is 
\begin{equation}
\Delta(\mathbf{r}) = \Delta_0 
\left[
\cos{\theta}\cos{(\mathbf{Q} \cdot \mathbf{r})} + \ii\sin{\theta}\sin{(\mathbf{Q} \cdot \mathbf{r})}
\right],
\label{eq:HFFLO}
\end{equation}
with $\Delta_0>0$ and $0\le\theta\le\pi/2$ (up to a gauge transformation) determined by the minimum of the free energy $\mathcal F=\langle \mathcal H\rangle$ at zero temperature.
The order parameter has a total magnitude
$|\Delta(\mathbf{r})|=\Delta_0\sqrt{(1+\cos{(2\theta)}\cos(2 \mathbf{Q}\cdot\mathbf{r}))/2}$,
having minima and maxima for any $\theta\neq\pi/4$ along the 1D planes parallel to the in-plane field, which we call respectively nodal and antinodal lines, satisfying $\mathbf{Q}\cdot\mathbf{r}=n \pi/2$ for $n\in\mathbb{Z}$.
Its phase $\phi(\mathbf{r})=\arg\Delta(\mathbf{r})$ is spatially modulated if $\theta\neq0,\pi/2$, being $\tan(\phi(\mathbf{r}))=\tan\theta\tan{(\mathbf{Q}\cdot\mathbf{r})}$ giving $\sin^2(\phi(\mathbf{r}))=0,1$ for $\mathbf{Q}\cdot\mathbf{r}=n \pi/2$.
One can verify that $\Delta(\mathbf{r},-\theta)=\Delta(\mathbf{r},\theta)^*$, $\Delta(\mathbf{r},\pi/2-\theta)=-\Delta(\mathbf{r},\pi/2+\theta)^*$, and that 
$\Delta(\mathbf{r},\pi/4-\theta)=\ii\Delta(\mathbf{r}',\pi/4+\theta)^*$ with $\mathbf{r}'={\pi\mathbf{Q}}/{2Q^2} - \mathbf{r}$.
Consequently, $H(\alpha+\theta)$ and $H(\alpha-\theta)$ are unitarily equivalent and thus have the same energy spectra, which mandates $\mathcal F (\alpha+\theta)=\mathcal F (\alpha-\theta)$ for $\alpha=0,\pi/4,\pi/2$.
This mandates the presence of stationary points $\delta\mathcal F(\theta)=0$ for $\theta=0,\pi/2$, and $\pi/4$ (see also Supplementary Note 3).
The cases $\theta=0,\pi/2$ correspond to Larkin–Ovchinnikov (LO) orders with a constant phase $\phi(\mathbf{r})=0,\pi/2$ and magnitude 
$\Delta_0|\cos{(\mathbf{Q}\cdot\mathbf{r}})|$ and 
$\Delta_0|\sin{(\mathbf{Q}\cdot\mathbf{r}})|$, respectively,
which becomes zero at the nodal lines and reaches its maximum $\Delta_0$ at the antinodal lines.
The case $\theta=\pi/4$ instead corresponds to a Fulde-Ferrel (FF) order with a constant magnitude $\Delta_0/\sqrt2$ and a phase $\phi(\mathbf{r})=\mathbf{Q}\cdot\mathbf{r}$ giving $\sin^2(\phi(\mathbf{r}))=0,1$ 
respectively for $\mathbf{Q}\cdot\mathbf{r}=n\pi$ and $\mathbf{Q}\cdot\mathbf{r}=\pi/2+n\pi$.
The 1D lines defined by constant $\mathbf{Q}\cdot\mathbf{r}$ have constant order parameter $\Delta=\Delta(\mathbf{r})$ and are described by an effective 1D Hamiltonian 
$H_\mathrm{1D}({r})=
\left[-v p_x \sigma_y\tau_z + 
(m' p_x^2 + m)\tau_x
\right]
\upsilon_z 
+\mathbf{b}\cdot\bm{\sigma}
-\Delta_0\,\tau_z\upsilon_y
|\Delta|
(
\cos\phi\,\upsilon_x+
\sin\phi\,\tau_z\upsilon_y
)
$,
for in-plane fields in the $x$ direction, which is equivalent to \cref{eq:HBdG} when one takes $p_y=0$.
In symmetry class D in 1D, topologically inequivalent phases are labeled by a $\nu\in\mathbb{Z}_2$ topological invariant. 
By dimensional reduction, $\nu$ must coincide with the parity of the topological invariant in 2D defined in \cref{eq:Z2invariant}:
Hence, there is only one nontrivial phase in 1D, realized when $\nu=1$ in \cref{eq:Z2invariant}, as long as the quasiparticle excitation gap remains open at all momenta (see Supplementary Figure 3).

\begin{figure*}[t]
\centering
\includegraphics[width=\textwidth]{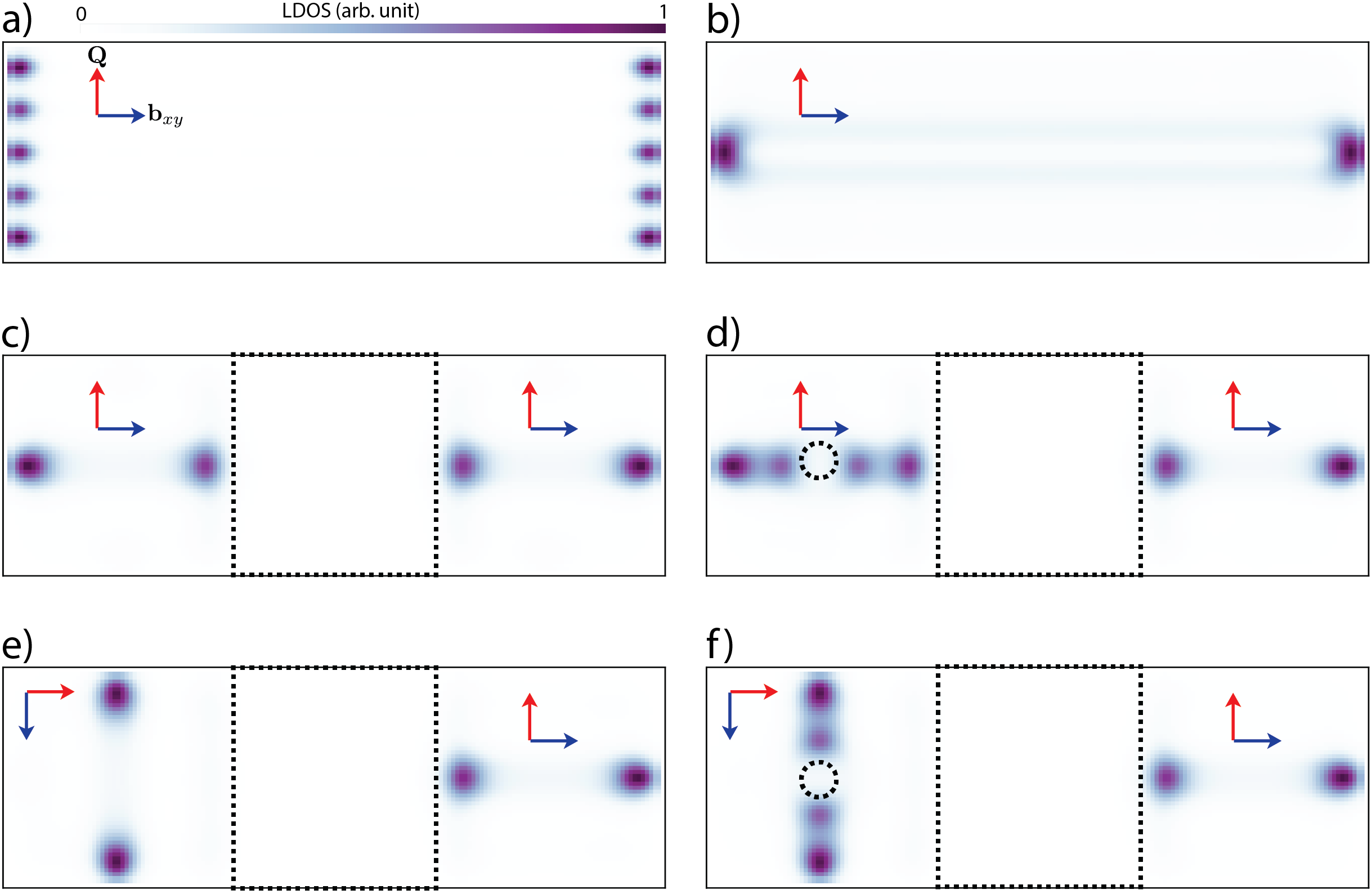} 
\caption{
Local density of states (LDOS) at zero energy calculated numerically for a proximitized topological insulator film in the inhomogeneous superconducting phase in a system of size $162\times 54$ lattice sites with open boundary conditions.
(a) Topologically nontrivial stripes corresponding to the order parameter phase $\phi(\mathbf{r})=\pm\pi/2$ with Majorana modes at their ends separated by a distance $\lambda/2=\pi/Q\approx \pi v/2{b}_{xy}$.
(b) A single stripe obtained by tuning the in-plane field such that $\lambda$ is approximately equal to the width of the system.
(c) same as (b), but with a trivial region in the middle, acting as a domain wall splitting the stripe into two. 
(d) same as (c), but with an additional trivial region on the left, acting as a pointlike defect further splitting the left stripe into two. 
(e) and (f) same as (c) and (d), but rotating the in-plane field on the left, (e) rotating the stripe, and (f) rotating two stripes around each other.
Arrows indicate the 
direction of the Cooper pair momentum $\mathbf{Q}$ and of the
in-plane field $\mathbf{b}_{xy}$. 
Areas enclosed by dotted lines indicate regions with $\mu=0$ suppressing the superconducting order.
The order parameter, calculated self-consistently, is compatible with $\theta=\pi/4$ in \cref{eq:HFFLO}.
}
\label{fig:ldos}
\end{figure*}

TNSs are realized when 1D lines $\mathbf{Q}\cdot\mathbf{r}=n \pi/2$ become topologically inequivalent.
For $\theta=0,\pi/2$, the order parameter phase is constant $\phi=0,\pi/2$:
for $\theta=\phi=0$, the gap closes at finite momenta
in the regime where $\nu=1$,
preventing the realization of a nontrivial gapped phase;
for $\theta=\phi=\pi/2$ instead, 
the gap is always open for $b_z\neq0$, and
TNSs may emerge when 
nontrivial phases are realized on the antinodal lines $\Delta(\mathbf{r})=\Delta_0$ for $b+|m|>\sqrt{\Delta_0^2+\mu^2}>|b-|m||$, and trivial phases on the nodal lines $\Delta(\mathbf{r})=0$ for $|b-|m||>|\mu|$ (nodal lines are $\Delta(\mathbf{r})=0$ and thus cannot realize a nontrivial TS). 
However, this phase is not physical since the superconductivity can only be realized when the Fermi level lies within the conduction band, i.e., for $|\mu|>b-|m|$.
For $\theta\approx\pi/4$ instead, the order parameter is $\Delta(\mathbf{r})\approx\Delta_0/\sqrt2$ which corresponds to trivial and nontrivial phases with $\phi(\mathbf{r})=\mathbf{Q}\cdot\mathbf{r}=0,\pi/2\mod\pi$, respectively, provided that 
\begin{equation}\label{eq:stripescondition}
2|m| \sqrt{\mu^2+{\Delta_0^2}/2}>
\vert m^2 + \mu^2 + {\Delta_0^2}/2 -b^2 \vert
>
2|m| |\mu|
,
\end{equation}
as it follows from \cref{eq:Z2invariant}.
The resulting TNSs are quasi-1D nontrivial regions close to the 1D lines $\mathbf{Q}\cdot\mathbf{r}=\pi n$ parallel to the in-plane field and effectively equivalent to 1D TSs. 
If stripes extend along the whole surface, reaching the hinges, there will be a MM at each end of the stripe, as in \cref{fig:system}(c).
These end modes can also be seen as the corner modes of the effectively 2D TSs obtained by extending the 1D lines along the $z$ direction, resulting in 2D planes parallel to the in-plane field and the $z$-axis, cutting the 3D TI into 2D slices.
Hence, the confinement of the 2D boundary Hamiltonian into a 1D Hamiltonian describing the TNSs corresponds to the confinement of the surface states of a 3D second-order topological phase (with hinge modes) into the edge states of a 2D second-order topological phase with corner modes defined by the planes with $\mathbf{Q}\cdot\mathbf{r}=\pi n$.

The formation of quasi-1D topological superconducting stripes and pointlike MMs at their ends is a consequence of dimensional reduction\cite{potter_multichannel_2010}.
The quasi-1D stipes are indeed narrow 2D regions which are topologically nontrivial,
with a length determined by the system size (or by the presence of domain walls) and a width $d<\lambda/2$ coinciding with the width of the region where the phase of the order parameter $\phi$ is such that $\nu=1$ in \cref{eq:Z2invariant}.
If their width is comparable with their length, these 2D nontrivial regions will exhibit 1D edge modes at their border on all four sides;
however, when their width becomes narrow enough, the edge states along two opposite sides will come closer and begin to overlap in space, opening a finite energy gap as a result of their finite overlap.
In particular, if the width is smaller or comparable to the Majorana localization length $d\lesssim\xi$, only a single quantization channel will become available.
In this regime, only a single pointlike mode may exist at each end of the stripe.
The dimensional reduction from a 2D to a 1D topological state requires stripes with a width smaller than their length and smaller than the Majorana localization length so that only one single 1D channel is present.
On the other hand, their spatial separation, given by the distance between neighboring stripes, must be larger than or comparable to the Majorana localization length $\lambda/2\gtrsim\xi$ so that MMs remain spatially separated.
Generally, one has $\xi\sim b/\Delta$ for 1D TSs\cite{klinovaja_composite_2012,mishmash_approaching_2016}.

As explained, symmetry arguments alone restrict the possible states to $\theta=0,\pi/2$ (LO states) and $\theta=\pi/4$ (FF state), but only the FF state with $\theta=\pi/4$ can exhibit TNSs.
We find numerical evidence that the state that fulfills the self-consistence equation at zero temperature has an order parameter which is approximately equal to \cref{eq:HFFLO} with $\theta=\pi/4$.
Indeed, we calculate the order parameter self-consistently at zero temperature and as a function of the spatial coordinate as 
$\Delta_{\tau}(\mathbf{r})
=
-U
\bra{\text{0}}
{\psi}_{\tau\down}(\mathbf{r})
{\psi}_{\tau\up}(\mathbf{r})
\ket{\text{0}}$ using \cref{eq:HFFLO} with several choices of $\theta$ and with $Q=2b_{xy}/v$ as the initial guess of the self-consistent calculation for realistic choices of the system parameters for \ce{Bi2Te3}\cite{chen_experimental_2009}, proximitized with \ce{NbTiN} or \ce{NbSe2}, compatible with \cref{eq:stripescondition}.
The resulting order parameter obtained self-consistently at zero temperature is approximately equal to \cref{eq:HFFLO} with $\theta=\pi/4$, corresponding to an FF order with almost constant magnitude and nonuniform phase, excluding regions close to the boundaries of the system, where the magnitude of the order parameter is slightly suppressed.
This result is in agreement with the results of Ref.~\onlinecite{hu_topological_2019}, which found that FFLO states with $\theta\approx\pi/4$ are stable also at finite temperature and for large in-plane magnetic fields $b_{xy}$
\footnote{
The ansatz for the order parameter in \cref{eq:HFFLO}, which describes a generic FFLO state interpolating between an FF state (for $\theta=0,\pi/2$) and an LO state (for $\theta=\pi/4$) coincides with the ansatz in Ref.~\cite{hu_topological_2019}, where the order parameter is parameterized in terms of $b=\cos(\theta)$. 
In Ref.~\cite{hu_topological_2019}, it is found $b=0.77$, which is approximately equal to $b=\cos(\theta)=\cos(\pi/4)=1/\sqrt{2}\approx0.707$, indicating an LO state (although, in that paper, the state is always called an FF state for any choice of $b$).
}.
Therefore, the superconducting order self-tunes to support the TNS phase, which is therefore a self-organized topological state, in this regard analogous to magnetic adatom chains with a spin helical order self-tuned to support the topological phase\cite{vazifeh_self-organized_2013}.
\Cref{fig:ldos}(a)
shows the local density of states (LDOS) at zero energy in the TNSs regime 
calculated numerically.
The peaks in the LDOS indicate MMs localized at the ends of the 1D TNSs at $\phi(\mathbf{r})=\pm\pi/2$.

\subsection*{Manipulation of topologically nontrivial stripes}

TNSs can be manipulated in several ways.
Rotating the magnetic field around the $z$-axis (perpendicular to the surface) changes the in-plane field direction and hence the stripes direction, while rotating the field in the $xy$-plane changes the in-plane field magnitude and hence the distance $\lambda/2$ between the stripes.
Moreover, topologically trivial regions $\nu=0$ can behave as domain walls or pointlike defects that confine the stripes or split a single stripe into two. 
Trivial regions can be created by locally increasing the tunneling $m$ between the two TI surfaces (by locally modifying the TI layer thickness), such that $|m|\gg|\mu|,|\Delta|,|b|$ or, alternatively, decreasing the chemical potential $|\mu|<b-|m|$ by using external gates, driving the Fermi level out of the conduction band and thus suppressing the superconducting order, so that the first term in \cref{eq:Z2invariant} dominates.
Trivial regions can also be created by suppressing the magnetic field since \cref{eq:Z2invariant} yields $\nu=0$ for $b=0$.
Furthermore, isolated stripes are obtained by tuning the distance $\lambda/2$ 
such that only a single stripe fits within the TNS phase (confined by the system edges or by domain walls), as in \cref{fig:ldos}(b).

\begin{figure*}[t]
\centering
\includegraphics[width=\textwidth]{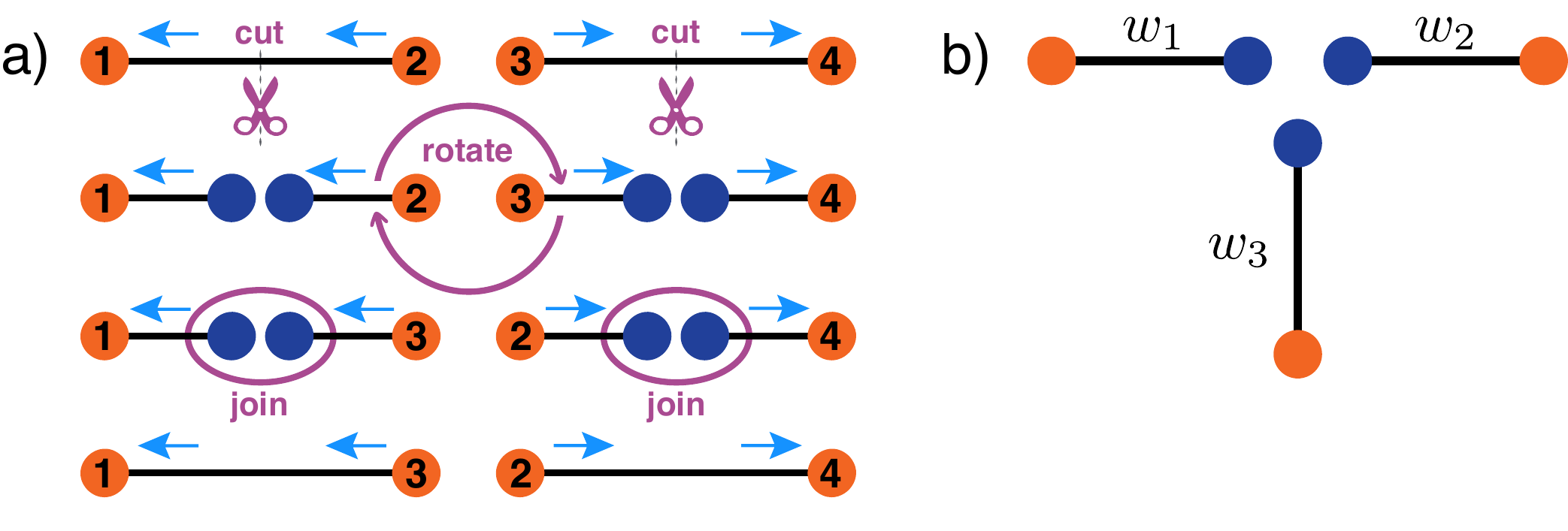} 
\caption{
Possible braiding protocols.
(a) Rotating left and right stripes exchange the modes $\gamma_1,\gamma_2$ and $\gamma_3,\gamma_4$, respectively.
Exchanging $\gamma_2,\gamma_3$ is obtained 
following a three steps protocol (from top to bottom):
i)
cut two stripes with antiparallel fields into four (creating four additional Majorana modes), 
ii)
rotate the resulting two inner stripes one around the other, following the arrows, leaving the other two outer stripes unchanged, and 
iii)
join the two inner stripes to the outer stripes.
Arrows indicate the direction of the in-plane field.
%$\mathbf{b}_{xy}$. 
(b) Braiding can also be performed in parameter space arranging three topologically nontrivial stripes around a pointlike defect and controlling the couplings $w_{1,2,3}$ between Majorana modes on opposite ends of the stripes, without moving them.
}
\label{fig:sketch}
\end{figure*}

\subsection*{Multilocational modes, SUSY, SYK model, and Yang-Lee anyons}

MMs at the ends of TNSs hybridize within a low-energy manifold of dimension $2N$, forming highly nonlocal modes delocalized over spatially separated points described by the effective Hamiltonian 
\begin{equation}
\mathcal{H}_\text{eff}=
\ii w \sum_{n=1}^{N-1} \sum_{s=L,R}
\gamma_{s,n} \gamma_{s,n+1}
+\ii w' \sum_{n=1}^N \gamma_{L,n} \gamma_{R,n},
\end{equation}
with $\gamma_{L,n}$ and $\gamma_{R,n}$
the modes on the left and right ends, 
$w\propto\ee^{-\lambda/2\xi}$ and $w'\propto\ee^{-l/\xi}$ the couplings between modes on the same side (at a distance $\lambda/2$) and on opposite sides of the stripes (at a distance $l$), respectively, with $w,w'>0$ up to a gauge transformation.
This manifold can exhibit nonlocal fermionic modes at exactly zero energy, even at finite sizes.
For $N\to\infty$ (or equivalently, in a setup with periodic boundaries), the MMs realize two translationally invariant lattices which are decoupled for $w'=0$.
In this case, the $N$ MMs of each lattice are degenerate under translations and hybridize into two MMs $\widetilde\gamma_{1,2}$ (forming a single fermionic mode) at zero energy, delocalized into $N$ spatially separated points corresponding to the ends of the TNSs.
This results in quantum mechanical SUSY\cite{hsieh_all-majorana_2016} or space-time SUSY in the presence of many-body interactions\cite{rahmani_phase_2015,rahmani_emergent_2015,sannomiya_supersymmetry_2019}.
For $w'>0$, the two fermionic modes (one for each side) hybridize at finite energy.
The fractionalization of the fermionic degrees of freedom and the emergence of quantum-mechanical SUSY is also observed in finite systems $N<\infty$ with open boundary conditions.
Indeed, we find that if $w=w'$ and $N=3m+2=2,5,8,\ldots$, the two nonlocal modes
\begin{align}
\widetilde\gamma_1\!=&\!\frac1{\sqrt{2m+2}}\!\sum_{n=0}^m\!
\kappa_{n}\!
(\!\gamma_{L\!,3n+1} \!+\! \gamma_{R\!,3n+2})
\!+\!
\kappa_{n-1}\!
(\!\gamma_{R\!,3n+1} \!-\! \gamma_{L\!,3n+2}),
%$
%and
%$
\\
\widetilde\gamma_2\!=&\!\frac1{\sqrt{2m+2}}\!\sum_{n=0}^m\!
\kappa_{n}\!
(\!\gamma_{R\!,3n+1} \!-\! \gamma_{L\!,3n+2})
\!+\!
\kappa_{n+1}\!
(\!\gamma_{L\!,3n+1} \!+\! \gamma_{R\!,3n+2}),
\end{align}
with $\kappa_n=\cos{\left({n\pi}/2\right)}$, have exactly zero energy, each delocalized into $2(m+1)=2(N+1)/3$ spatially separated points.
This regime $w'=w$ is realized when the length of the stripes is comparable with the distance between mutual stripes (see Supplementary Note 6).
This setup can be advantageous when the system dimensions cannot be stretched indefinitely.
Also, we find that if $w'\to0$ and $N=2m+1$, the 
two nonlocal modes 
\begin{align}
%$
\widetilde\gamma_1=\frac1{\sqrt{m+1}}\sum_{n=0}^{m} \gamma_{L,2n+1},
\qquad
%$, 
%and 
%$
\widetilde\gamma_2=\frac1{\sqrt{m+1}}\sum_{n=0}^{m} \gamma_{R,2n+1},
%$
\end{align}
have exactly zero energy, each delocalized into $m+1=(N+1)/2$ spatially separated points at the ends of every other stripe.
This case is realized in \cref{fig:ldos}(a) since $w'\approx0$ (see Supplementary Figure 4).
The case $w'=0$ is realized asymptotically when the stripes become infinitely long.
Hence, this case requires one of the system dimensions to be much larger than the other.
These two cases are further examples of multi-locational MMs recently predicted to appear in three-terminal Josephson junctions\cite{nagae_multilocational_2024}.
In all these cases, the groundstate is twofold degenerate, with two nonlocal MMs $\widetilde\gamma_1$ and $\widetilde\gamma_2$ forming a single zero-energy and particle-hole symmetric fermionic mode.
One can thus define two fermionic operators $Q_{1,2}=\widetilde\gamma_{1,2}\sqrt{H_\text{SUSY}}$ satisfying the algebra $\{P,Q_i\}=0$, $\{Q_i,Q_j\}=2\delta_{ij}H_\text{SUSY}$, where $H_\text{SUSY}$ is the many-body Hamiltonian with all energy levels positive (obtained by adding a positive constant) and $P$ the fermion parity operator.
This corresponds to spontaneously broken ${\cal N}=2$ quantum mechanical SUSY\cite{witten_dynamical_1981} with supercharges $Q_{1,2}$, zero superpotential, and Witten index $W=1$. 
Furthermore, configurations with several stripes, as in \cref{fig:ldos}(a), are equivalent to sets of equally-spaced 1D TSs, which can effectively realize the SYK model when coupled to a quantum dot\cite{chew_approximating_2017}, or Yang-Lee anyons when coupled to a metallic bath\cite{sanno_engineering_2022}, provided that the single-particle couplings $w$ and $w'$ are suppressed\cite{garcia-garcia_chaotic-integrable_2018}.
Finally, note that for $N\to\infty$, the two MMs chains at opposite edges form a pair of 1D chiral Majorana edge modes with finite dispersion\cite{zhang_topological_2013,hu_topological_2019,marra_1d-majorana_2022}.

\subsection*{Braiding}

MMs at the ends of TNSs can be braided.
To do so, we need the ability to
i) rotate stripes to exchange the MMs and 
ii) split or merge stripes to implement fusion and read-out.
TNSs can be rotated by rotating the in-plane field.
Adiabatically decreasing the chemical potential (thus suppressing superconductivity) on specified regions can create domain walls or pointlike defects that split single stripes into two stripes with parallel magnetic fields. 
The reverse process merges two stripes with parallel fields into one.
Figures~\ref{fig:ldos}(c) and~\ref{fig:ldos}(d) show how to adiabatically split a single stripe into two segments via a domain wall in the middle, and rotate one segment by rotating the in-plane field in one half of the system.
Figures~\ref{fig:ldos}(e) and~\ref{fig:ldos}(f) show how to rotate two stripes around a central pointlike defect.
The lowest energy levels corresponding to the MMs remained close to zero in all cases (see Supplementary Figures 6 and 7).
\Cref{fig:sketch}(a) illustrates a possible braiding protocol.
Additionally, TNSs can be controlled by moving the domain walls.
Alternatively, braiding can be performed in parameter space\cite{sau_controlling_2011,karzig_shortcuts_2015} without moving the TNSs, e.g., arranging three TNSs around a pointlike defect as in \cref{fig:sketch}(b) and controlling the coupling between MMs on opposite ends of the TNSs (see Supplementary Figure 8).

\section*{Discussion}

In this work, we described an alternative 2D platform to create, manipulate, and braid MMs via inhomogeneous superconducting orders in proximitized TIs.
Unlike other 2D platforms, MMs do not localize at the vortex cores of the order parameter but at the opposite ends of TNSs induced by the inhomogeneous order.
This setup can realize topological quantum gates and other exotic quantum phenomena, such as quantum mechanical SUSY, Yang-Lee anyons, and the SYK model.
Moreover, TNSs may also be induced by inhomogeneous superconducting orders
in \ce{Sr2RuO4}\cite{kinjo_superconducting_2022}, iron pnictides\cite{kasahara_evidence_2020},
organic superconductors\cite{coniglio_superconducting_2011,sugiura_fuldeferrelllarkinovchinnikov_2019,sari_distorted_2021},
\ce{SrTiO3}/\ce{LaAlO3} interfaces\cite{michaeli_superconducting_2012},
\ce{KTaO3}/\ce{EuO} or \ce{KTaO3}/\ce{LaAlO3} interfaces in the inhomogeneous superconducting stripe phase\cite{liu_two-dimensional_2021}, and 
two-component cold atomic Fermi gases with population imbalance and effective spin-orbit coupling\cite{zwierlein_fermionic_2006,takahashi_vortex-core_2006,suzuki_magnetization_2008,zhang_topological_2013,qu_topological_2013,inotani_radial_2021}.
Finally, the experimental detection of TNSs would also provide indirect evidence of FFLO inhomogeneous superconductivity.

\section*{Methods}

The numerical results were obtained by discretizing the continuous Hamiltonian into a lattice model and calculating the energy spectra, wavefunction, and superconducting order parameter self-consistently at zero temperature.
The local density of states (LDOS) at zero energy was calculated directly from the energy spectra and wavefunction.
The parameters used for the numerical calculations were chosen to be compatible with heterostructures of
\ce{Bi2Te3}\cite{chen_experimental_2009} proximitized with \ce{NbTiN} or \ce{NbSe2}.
The topological invariants (Chern numbers) were calculated numerically using the Fukui-Hatsugai-Suzuki method\cite{fukui_chern_2005}, while the parity of the topological invariants was calculated as the sign of the product of the Pfaffians of the Hamiltonian in the Majorana basis with momenta spanning the time-reversal symmetry points in the Brillouin zone.

\section*{Data availability}

The code used for the numerical simulations within this paper and the resulting data are available from the corresponding author on reasonable request.

\begin{acknowledgments}
P.~M. thanks Paolo Mele, Dita Puspita Sari, Masatoshi Sato, and Ken Shiozaki for useful discussions, and Maria Bartho for help preparing figures.
P.~M. is supported by the Japan Science and Technology Agency (JST) of the Ministry of Education, Culture, Sports, Science and Technology (MEXT), JST CREST Grant~No.~JPMJCR19T2, the Japan Society for the Promotion of Science (JSPS) Grant-in-Aid for Early-Career Scientists KAKENHI Grant~No.~JP23K13028 and No.~JP20K14375.
T.~M. is supported by the Grant-in-Aid for Scientific Research on Innovative Areas ``Quantum Liquid Crystals’' Grant No.~JP22H04480 and JSPS KAKENHI Grant No.~JP20K03860, No.~JP21H01039, and No.~JP22H01221.
M.~N. is supported by JSPS KAKENHI Grant No.~JP22H01221 and by the WPI program ``Sustainability with Knotted Chiral Meta Matter (SKCM2)'' at Hiroshima University.
\end{acknowledgments}

\section*{Author contributions}

All authors P.~M., D.~I., T.~M., and M.~N. contributed to the scientific discussion and to the final revision of the manuscript.
P.~M. carried out the numerical calculations and wrote the initial draft.

\section*{Competing interests}

The authors declare no competing interests.

\end{document}

\title{
Majorana modes in striped two-dimensional inhomogeneous topological superconductors:
\\
Supplementary Material
}
\author{Pasquale Marra}
\email{pmarra@ms.u-tokyo.ac.jp}
\affiliation{
Graduate School of Mathematical Sciences,
The University of Tokyo, 3-8-1 Komaba, Meguro, Tokyo, 153-8914, Japan
}
\affiliation{
Department of Physics \& Research and Education Center for Natural Sciences, 
Keio University, 4-1-1 Hiyoshi, Yokohama, Kanagawa, 223-8521, Japan
}
\author{Daisuke Inotani} 
\affiliation{
Department of Physics \& Research and Education Center for Natural Sciences, 
Keio University, 4-1-1 Hiyoshi, Yokohama, Kanagawa, 223-8521, Japan
}
\author{Takeshi Mizushima} 
\affiliation{
Department of Materials Engineering Science, Osaka University, Toyonaka, Osaka 560-8531, Japan
}
\author{Muneto Nitta} 
\affiliation{
Department of Physics \& Research and Education Center for Natural Sciences, 
Keio University, 4-1-1 Hiyoshi, Yokohama, Kanagawa, 223-8521, Japan
}
\affiliation{
International Institute for Sustainability with Knotted Chiral Meta Matter (SKCM$^2$), Hiroshima University, 1-3-2 Kagamiyama, Higashi-Hiroshima, Hiroshima, 739-8511, Japan
}
\begin{abstract}
Here, we provide additional details on 
the bare electron dispersion (Supplementary Note 1), 
the calculation of the topological invariants (Supplementary Note 2), 
the Fulde-Ferrel-Larkin–Ovchinnikov (FFLO) order (Supplementary Note 3), 
the self-consistent numerical calculations of the order parameter and local density of states (Supplementary Note 4), 
alternative braiding protocols (Supplementary Note 5),
and the existence of zero energy superpositions of Majorana modes localized at the ends of topologically nontrivial stripes (Supplementary Note 6).
\end{abstract}
\maketitle
\date\today

\setcounter{section}{0}
\setcounter{equation}{0}

\section*{Supplementary Note 1. Bare electron dispersion}

\begin{figure}[t]
\centering
\includegraphics[width=0.5\columnwidth]{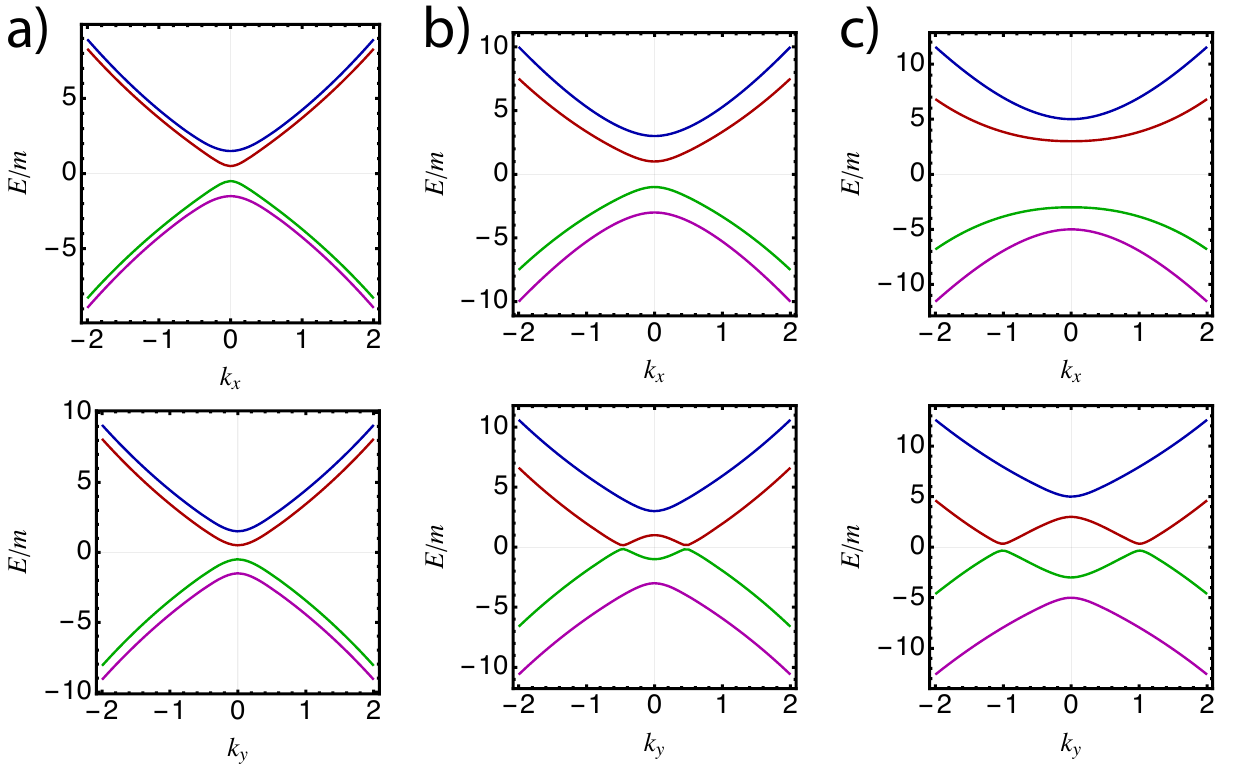} 
\caption{
Energy dispersion in different regimes phases in the normal state with for different choices of the magnetic field $b$, with in-plane field parallel to the $x$ axis and a small out-of-plane component (i.e., $b_x,b_z\neq0$, $b_y=0$) 
and with $m'=m$, $\mu=0$, $v=3.3m$.
(a) 
Energy dispersion with $b=0.5m$.
(b)
Energy dispersion with $b=2m$.
(c)
Energy dispersion with $b=4m$.
}
\label{fig:baredispersion}
\end{figure}

In the normal regime $\Delta=0$ with broken time-reversal symmetry (finite fields $b>0$), the boundary Hamiltonian of the topological insulator surface states is given by 
\begin{equation}
H = (m' \mathbf{p}^2 + m) \tau_x + v(p_y \sigma_x - p_x \sigma_y)\tau_z - \mu
+ \mathbf{b}\cdot\bm{\sigma},
\end{equation}
with a bare energy dispersion with four branches 
\begin{equation}
E(\mathbf{k})=\pm\sqrt{(m' k^2 + m)^2 + v^2 k^2 + b^2 \pm 2 \sqrt{b^2 (m' k^2 + m)^2 + v^2(b_y k_x - b_x k_y)^2}}-\mu,
\end{equation}
with an energy gap at $k=0$ equal to $2| b - |m| |$.
If the out-of-plane field vanishes $b_z=0$ and for $b>|m|$, the gap closes at finite momenta $k^2={(\sqrt{4 b^2 m'^2+ 4m m' v^2+v^4}-2 m m'-v^2)}/{2 m'^2}$ perpendicular to the field.
For $m',m,b_z\ll b_{xy}$, the dispersion has a minimum for momenta $|k|=b_{xy}/v$ and with momenta $\mathbf{k}$ perpendicular to the in-plane field $\mathbf{b}_{xy}$ 
(e.g., $k_x=\pm b_{y}/v$ for in-plane field in the $y$ direction,
and $k_y=\pm b_{x}/v$ for in-plane field in the $x$ direction).
The energy dispersion for some choices of the magnetic field is shown in \cref{fig:baredispersion}.

\section*{Supplementary Note 2. Topological phases in the superconducting regime with broken time-reversal symmetry}

Let us discuss the topological phases in the superconducting regime with broken time-reversal symmetry (finite fields $b>0$) and uniform superconducting pairing $\Delta>0$. 
The Bogoliubov-de~Gennes (BdG) Hamiltonian density 
describing the system 
is
\begin{equation}
H_\mathrm{BdG}=
\left[
( m' \mathbf{p}^2 + m)\tau_x
-v\mathbf{p}\times\bm{\sigma}\tau_z
-\mu
\right]
\upsilon_z 
+
\mathbf{b}\cdot\bm{\sigma}
+\Delta\cos\phi\,\upsilon_x
-\Delta\sin\phi\,\tau_z\upsilon_y,
\label{eq:HBdG}
\end{equation}
with the energy gap closing at $k=0$ if $|m^2+\mu^2+\Delta^2-b^2| =2 |m| \sqrt{\mu^2+\Delta ^2 \sin^2\phi}$, specifically, being equal to $2\min(|b\pm\sqrt{\Delta ^2+(m\pm\mu)^2}|)$ and $2\min(| |m|-| b\pm\sqrt{\Delta^2+\mu^2}| | )$ for $\phi=0,\pi/2$, respectively.
If the magnetic field is in-plane (i.e., $b_z=0$), the energy gap closes at finite momenta $k>0$ for large enough fields $|b_{xy}|\gtrsim\Delta$ for any choice of the phase $\phi$.
This is because, in this case, electrons with momenta perpendicular to the in-plane field becomes spin-polarized in the direction of the in-plane fields.
Since a uniform $s$-wave superconducting pairing term cannot couple electrons with same spin and opposite momenta, the pairing vanishes and the 
quasiparticle excitation gap 
 closes.
Moreover, for $\phi=0$ the gap closes at finite momenta $k>0$ at large fields, regardless the direction of the field.
Therefore, we will assume hereafter that $b_z>0$ to keep the energy gap open at large momenta in all regimes.

The BdG Hamiltonian exhibits particle-hole symmetry (PHS) defined as $\mathcal{C} H_\mathrm{BdG} \mathcal{C}^{-1}=-H_\mathrm{BdG}$, where $\mathcal{C}=\sigma_y\upsilon_y\mathcal{K}$ with $\mathcal{C}^2=1$.
However, at finite magnetic fields, the Hamiltonian exhibits broken time-reversal symmetry (TRS), 
Hence, the Hamiltonian belongs to the symmetry class D in 2D and can exhibit topologically nontrivial phases labelled by a topological invariant $c\in\mathbb{Z}$, which coincides with the Chern number of the 
quasiparticle excitation gap.
The parity of the topological invariant is related to the pfaffian invariant\cite{tewari_topological_2012a,budich_equivalent_2013} $(-1)^{c}=\prod_{\mathbf{k}}\sgn\left(\pf\left(H_\mathrm{BdG}(\mathbf{k})\sigma_y\upsilon_y\right)\right)$
where the product span over the TRS points $\mathbf{k}=(0,0)$, $(0,\infty)$, $(\infty,0)$, $(\infty,\infty)$ in the continuum, or equivalently $\mathbf{k}=(0,0)$, $(0,\pi)$, $(\pi,0)$, $(\pi,\pi)$, by regularizing the Hamiltonian on a discrete lattice.
This can be obtained directly by taking $k\to\sin k$ and $k^2\to2-2\cos k$ (in units of the lattice parameter $a=1$), which yields
\begin{align}
H_\mathrm{BdG}(\mathbf{k})=&
\left[v(\sin{k_y} \sigma_x - \sin{k_x} \sigma_y)\tau_z + 
( m' (4 - 2\cos{k_x} - 2\cos{k_y}) + m)\tau_x
-\mu
\right]
\upsilon_z +
\mathbf{b}\cdot\bm{\sigma}
\nonumber\\&
+\Delta\cos\phi\,\upsilon_x
-\Delta\sin\phi\,\tau_z\upsilon_y.
\end{align}
Due to rotation symmetry, the pfaffian assumes the same value at $\mathbf{k}=(0,\pi), (\pi,0)$ and thus the product can be restricted to the remaining two TRS points $\mathbf{k}=(0,0), (\pi,\pi)$, which yields
\begin{align}
(-1)^{c}=&\sgn
\left(\left(m^2+\Delta^2+\mu^2-b^2\right)^2-4 m^2 \left(\mu^2+\Delta^2 \sin^2\phi\right)\right)
\nonumber\\
\times 
&\sgn
\left(\left((m+8 m')^2+\Delta^2+\mu^2-b^2\right)^2-4 (m+8 m')^2 \left(\mu^2+\Delta^2 \sin ^2\phi\right)\right).
\end{align}
The second factor in the equation above, which correspond to the TRS point $\mathbf{k}=(\pi,\pi)$, gives
\begin{gather}
\sgn\left(\left((m+8 m')^2+\Delta^2+\mu^2-b^2\right)^2-4 (m+8 m')^2 \left(\mu^2+\Delta^2 \sin ^2\phi\right)\right)
\ge
\nonumber\\
\sgn\left(\left((m+8 m')^2+\Delta^2+\mu^2-b^2\right)^2-4 (m+8 m')^2 \left(\mu^2+\Delta^2\right)\right)
=\nonumber\\
\sgn\left(
\left(|m+8 m' -b| -\sqrt{\Delta^2+\mu^2}\right) 
\left(|m+8 m' +b| -\sqrt{\Delta^2+\mu^2}\right) 
\right),
\end{gather}
which is positive when $\min(|m+8 m' \pm b|) > \sqrt{\Delta^2+\mu^2}$.
Hence, when $|m'|\gg b,\Delta,|m|,|\mu|$ is the dominant energy scale, one obtains
\begin{equation}
(-1)^{c}=\sgn
\left(\left(m^2+\Delta^2+\mu^2-b^2\right)^2-4 m^2 \left(\mu^2+\Delta^2 \sin^2\phi\right)\right),
\end{equation}
which gives Eq.~(4) of the main text.

\subsection{Case $\phi=\pi/2$}

\begin{figure}[t]
\centering
\includegraphics[width=\columnwidth]{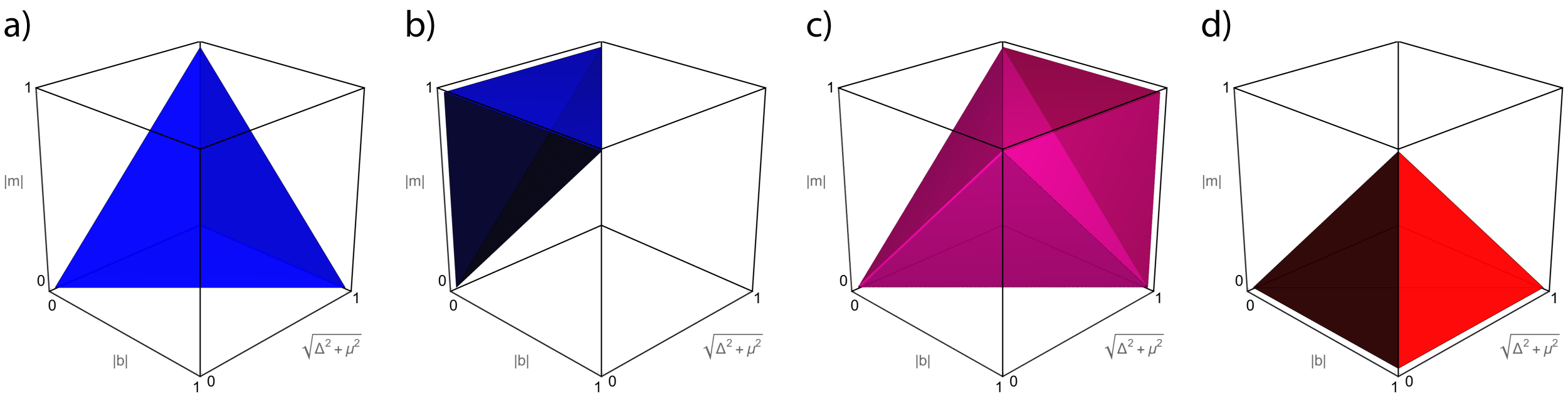} 
\caption{
Topological phases in the homogeneous superconducting order with $\phi=\pi/2$ as a function of $b$,$m$, and $\sqrt{\Delta^2+\mu^2}$.
(a) and (b)
Phase at weak fields realized for 
$\sqrt{\Delta^2+\mu^2}>b+|m|$
and for 
$|m|>b+\sqrt{\Delta^2+\mu^2}$.
This phase is topologically trivial, with Chern number $c=0$ and parity $\nu=0$.
(c)
Phase at intermediate fields,
realized for 
$|m|+\sqrt{\Delta^2+\mu^2}>b>| |m|-\sqrt{\Delta^2+\mu^2} |$, 
or equivalently, $|m|+b>\sqrt{\Delta^2+\mu^2}>| |m|-b |$, 
or $b+\sqrt{\Delta^2+\mu^2}>|m|>| |b-\sqrt{\Delta^2+\mu^2} |$.
This phase is topologically nontrivial with Chern number $c=\pm1$ and parity $\nu=1$.
(d)
Phase at strong fields realized for $b>|m|+\sqrt{\Delta^2+\mu^2}$.
This phase is topologically nontrivial with Chern number $c=\pm2$ and parity $\nu=0$.
}
\label{fig:phasediagram}
\end{figure}

In the case that $\phi=\pi/2$, the 
quasiparticle excitation gap 
 $2\min(| |m|-| b\pm\sqrt{\Delta^2+\mu^2}| | )$ at $k=0$ closes when any of the quantities $b \pm m \pm\sqrt{\Delta^2+\mu^2}$ is zero, with the parity of the topological invariant given by
\begin{equation}
(-1)^{c}=\sgn
\left(\left(m^2+\Delta^2+\mu^2-b^2\right)^2-4 m^2 \left(\mu^2+\Delta^2\right)\right)
=
\sgn
\prod_{\pm\pm}
(b\pm m\pm\sqrt{\Delta^2+\mu^2}) 
,
\end{equation}
where the product span over all four possible combinations of the signs $\pm$.
Hence, there is a quantum phase transition each time that the quantities $b\pm m\pm\sqrt{\Delta^2+\mu^2}$ changes its sign.
This condition divides the parameter space into 4 distinct 
regions 
separated by the closing of the 
quasiparticle excitation gap:
A 
region 
at strong field $b>|m|+\sqrt{\Delta^2+\mu^2}$, 
two 
regions 
at weak field $\sqrt{\Delta^2+\mu^2}>b+|m|$ and $|m|>b+\sqrt{\Delta^2+\mu^2}$, and 
an intermediate 
region 
where no energy scales dominates, i.e., when 
$|m|+\sqrt{\Delta^2+\mu^2}>b>| |m|-\sqrt{\Delta^2+\mu^2} |$, 
or equivalently, $|m|+b>\sqrt{\Delta^2+\mu^2}>| |m|-b |$, 
or $b+\sqrt{\Delta^2+\mu^2}>|m|>| |b-\sqrt{\Delta^2+\mu^2} |$.
In the strong and weak field 
regions 
one has $c\equiv0\mod2$, while in the intermediate 
region 
one has $c\equiv1\mod2$.
The 
two regions 
at weak fields are 
in the same phase, which is topologically
trivial with $c=0$, since they can be continuously connected with the phase at zero field $b\to0$ with TRS, which is trivial in class D.
The phase at strong field and the intermediate phase need to have even and odd topological invariant $c\equiv0$, $1\mod2$, respectively.
The different regions and phases as a function of $b$ ,$m$, and $\sqrt{\Delta^2+\mu^2}$ for $\phi=\pi/2$ are shown in \cref{fig:phasediagram}.
The energy spectra is always gapped in the weak fields regime.
In the intermediate and strong fields regimes, the energy spectra is always gapped for $b_z\neq0$.
The corresponding energy dispersion in the different cases is shown in \cref{fig:dispersion}.

To determine the value of the topological invariant we calculate numerically\cite{fukui_chern_2005} the Chern number at half-filling by sampling the parameter space by different combinations of the parameters $b_z,m,\Delta$ and taking $m'=v=1$, $\mu=0$, and $b_{xy}=0$. 
For the phase at weak fields we take $b_z=0$, $\Delta=1$, and $m=\pm 1/2$, or alternatively, $b_z=0$, $\Delta=0$, and $m=\pm 1/2$, and obtain $c=0$ in all cases, as expected.
For the phase at strong field we take $b_z=\pm 1$, $\Delta=0$, and $m=\pm 1/2$, and obtain $|c|=2$.
For the intermediate phase we take and $m=\pm 1/2$, and obtain $|c|=1$.
Since the Chern number does not change up to continuous transformations which keeps the gap open, we conclude that 
$c=\pm2$ for the phase at strong field,
$c=0$ for the phase at weak fields, and $c=\pm1$ for the intermediate phase.
The nontrivial phases are still realized for $\phi\neq\pi/2$ as long as the 
quasiparticle excitation gap 
 does not close at $k=0$ for
\begin{equation}
\sin^2\phi^\prime=
\frac{(m^2+\mu^2+\Delta^2-b^2)^2 - 4m^2\mu^2 }{4m^2\Delta ^2},
\end{equation}
or at finite momenta $k>0$.

\subsection{Case $\phi=0$}

In the case that $\phi=0$, the 
quasiparticle excitation gap 
 $2\min(|b-\sqrt{\Delta ^2+(m\pm\mu)^2}|)$ at $k=0$ closes when any of the quantities $b\pm\sqrt{\Delta ^2+(m\pm\mu)^2}|)$ is zero, with the parity of the topological invariant given by
\begin{equation}
(-1)^{c}=\sgn
\left(b^2-\Delta^2-(m-\mu)^2\right)\left(b^2-\Delta^2-(m+\mu)^2\right)=
\sgn\prod_{\pm\pm}
\left(b\pm\sqrt{\Delta^2+(m\pm\mu)^2}\right)
,
\end{equation}
where the product spans over all four possible combinations of the signs $\pm$.
Hence, there is a phase transition each time that the quantities $b-\sqrt{\Delta^2+(m\pm\mu)^2}$ changes its sign.
This condition divides the parameter space into 3 distinct phases:
A phase at strong field $b>\sqrt{\Delta^2+(m\pm\mu)^2}$, 
a phase at weak field $b<\sqrt{\Delta^2+(m\pm\mu)^2}$, and
an intermediate phase $\sqrt{\Delta^2+(|m|+|\mu|)^2}>b>\sqrt{\Delta^2+(|m|-|\mu|)^2}$.
The phases at weak field is trivial with $c=0$, since they can be continuously connected with the phase at zero field $b\to0$ with TRS, which is trivial in class D.

The intermediate phase is not gapped, since the 
quasiparticle excitation gap 
 closes at small momenta.
At $k=0$ the Hamiltonian can be diagonalized with respect to the particle-hole and top-bottom sectors, and written as
$
H_\mathrm{BdG}=
(
m\tau_z
-\mu
)
\upsilon_x
+
\mathbf{b}\cdot\bm{\sigma}
+\Delta\,\upsilon_z
$,
and therefore at small momenta, by neglecting terms in $p^2$, the Hamiltonian becomes 
\begin{equation}
H_\mathrm{BdG}=
(
m\tau_z
-\mu
)
\upsilon_x
+
\mathbf{b}\cdot\bm{\sigma}
+\Delta\,\upsilon_z
-v\mathbf{p}\times\bm{\sigma}\tau_x\upsilon_x
,
\end{equation}
which can be recast into two energy sectors
\begin{equation}
H_\pm=
(
m\pm\mu
)
\upsilon_x
+
\mathbf{b}\cdot\bm{\sigma}
+\Delta\,\upsilon_z
,
\end{equation}
coupled at small momenta by terms $\propto v\mathbf{p}\times\bm{\sigma}$.
In the weak field phase $b<\sqrt{\Delta^2+(m\pm\mu)^2}$, assuming $|m-\mu|<|m+\mu|$, the lowest energy levels below and above the gap are $b-\sqrt{\Delta ^2+(m-\mu )^2}<0$, $-b+\sqrt{\Delta ^2+(m-\mu )^2}>0$, corresponding to the energy sector $H_-$.
The transition to the intermediate phase $b>\sqrt{\Delta^2+(m\pm\mu)^2}$, correspond to the inversion of the gap, with the two energy levels becoming now $b-\sqrt{\Delta^2+(|m|-|\mu|)^2}>0$, $-b+\sqrt{\Delta ^2+(|m|-|\mu|)^2}<0$.
However, the corresponding energy bands of these two energy levels are not gapped at finite momenta, since they are not coupled by terms at finite momenta $\propto v\mathbf{p}\times\bm{\sigma}$, and therefore they become degenerate and cross zero energy at some finite momentum $k>0$.

\begin{figure}[t]
\centering
\includegraphics[width=\columnwidth]{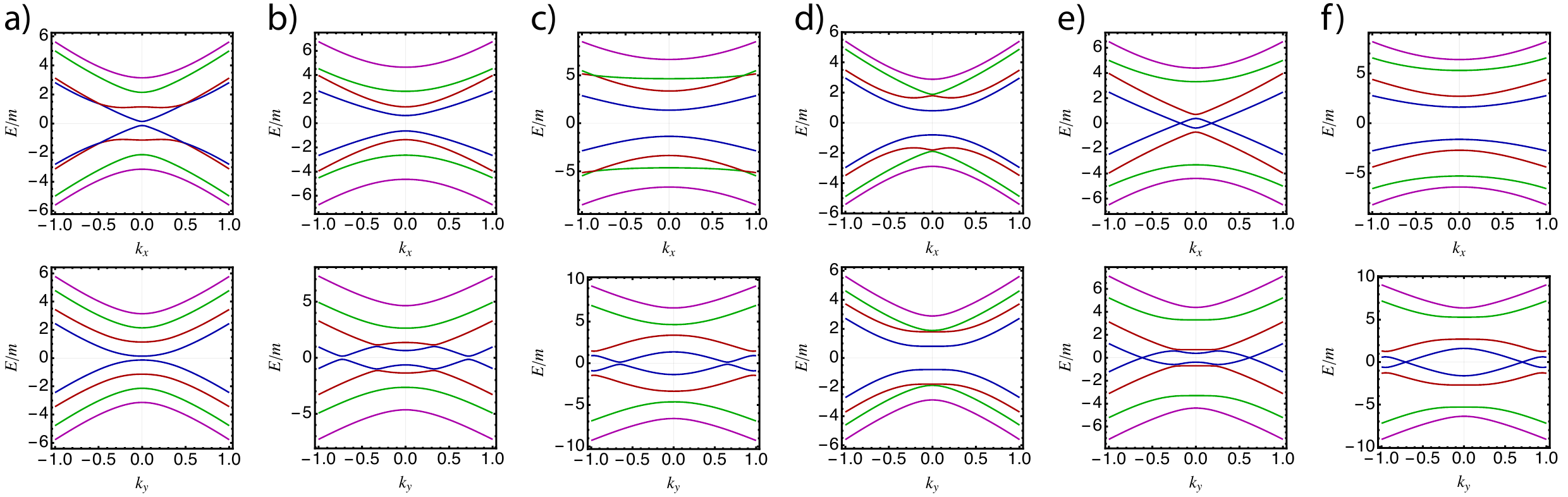} 
\caption{
Energy dispersion in different regimes phases in the homogeneous superconducting order for $\phi=\pi/2$ and $\phi=0$ and different choices of the magnetic field $b$, with in-plane field parallel to the $x$ axis and a small out-of-plane component (i.e., $b_x,b_z\neq0$, $b_y=0$) 
and with $m'=m=\mu$, $v=3.3m$, $\Delta=1.3m$.
(a) 
Energy dispersion with $\phi=\pi/2$ and $b=0.5m$
corresponding to the phase at weak fields 
$\sqrt{\Delta^2+\mu^2}>b+|m|$.
(b)
Energy dispersion with $\phi=\pi/2$ and $b=2m$
corresponding to the phase at intermediate fields
$|m|+\sqrt{\Delta^2+\mu^2}>b>| |m|-\sqrt{\Delta^2+\mu^2} |$.
This phase becomes gapless at finite momenta for $b_z=0$ (not shown).
(c)
Energy dispersion with $\phi=\pi/2$ and $b=4m$
corresponding to the phase at strong fields
$b>|m|+\sqrt{\Delta^2+\mu^2}$.
Also this phase becomes gapless at finite momenta for $b_z=0$ (not shown).
(d)
Energy dispersion with $\phi=0$ and $b=0.5m$
corresponding to the phase at weak fields 
$b<\sqrt{\Delta^2+(m\pm\mu)^2}$.
(e)
Energy dispersion with $\phi=0$ and $b=2m$
corresponding to the phase at intermediate fields
$\sqrt{\Delta^2+(|m|+|\mu|)^2}>b>\sqrt{\Delta^2+(|m|-|\mu|)^2}$.
Notice that in this phase, the energy spectra is not gapped at finite momenta.
(f)
Energy dispersion with $\phi=0$ and $b=4m$
corresponding to the phase at strong fields
$b>\sqrt{\Delta^2+(m\pm\mu)^2}$.
In this phase as well, the energy spectra is not gapped at finite momenta.
This phase becomes fully gapped if the in-plane field is zero (i.e., $b_x=b_y=0$ and $b_z=b$).
}
\label{fig:dispersion}
\end{figure}

In the strong field regime, the Chern invariant is even $c\equiv0\mod2$.
To determine the value of the topological invariant for the phase at strong field we calculate numerically\cite{fukui_chern_2005} the Chern number at half-filling with $m'=v=1$, $\mu=0$, $b_{xy}=0$, $b_z=\pm 1$, $\Delta=0$, and $m=\pm 1/2$, and obtain $|c|=2$.
Also in this regime, the energy dispersion is not gapped at finite momenta and finite fields unless $b_x=b_y=0$ and $b_z\neq0$.
The energy dispersion in the different cases is shown in \cref{fig:dispersion}.

\subsection{Topological phases in the effectively 1D superconducting regime}

In the effectively 1D regime induced by the LO state, the topological invariant along the antinodal and nodal lines (and any line parallel to them) is $\nu\in\mathbb{Z}_2$ in class D, being $(-1)^{\nu}=\prod_{\mathbf{k}}\sgn\left(\pf\left(H_\mathrm{BdG}(k_x,0)\sigma_y\upsilon_y\right)\right)$ where the product span the TRS points $k_x=0, \pi$, which yields
\begin{align}
(-1)^{\nu}=&\sgn
\left(\left(m^2+\Delta^2+\mu^2-b^2\right)^2-4 m^2 \left(\mu^2+\Delta^2 \sin^2\phi\right)\right)
\nonumber\\
\times 
&\sgn
\left(\left((m+4 m')^2+\Delta^2+\mu^2-b^2\right)^2-4 (m+4 m')^2 \left(\mu^2+\Delta^2 \sin ^2\phi\right)\right).
\end{align}
The second factor in the equation above, which corresponds to the TRS point $\mathbf{k}=(\pi,0)$, is positive when $\min(|m+4 m' \pm b|) > \sqrt{\Delta^2+\mu^2}$.
Therefore, when $|m'|\gg b,|m|,|\mu|,\Delta$ is the dominant energy scale, one obtains again
\begin{equation}
(-1)^{\nu}=\sgn
\left(\left(m^2+\Delta^2+\mu^2-b^2\right)^2-4 m^2 \left(\mu^2+\Delta^2 \sin^2\phi\right)\right).
\end{equation}
Hence, the trivial and nontrivial phases $c=0,\pm2$ in 2D correspond to the trivial phase $\nu=0$ in 1D, while the nontrivial phases with $c=\pm1$ in 2D correspond to the nontrivial phase $\nu=1$ in 1D.

\section*{Supplementary Note 3. FFLO order}

\begin{figure}[t]
 \centering
 	\includegraphics[height=.6\columnwidth]{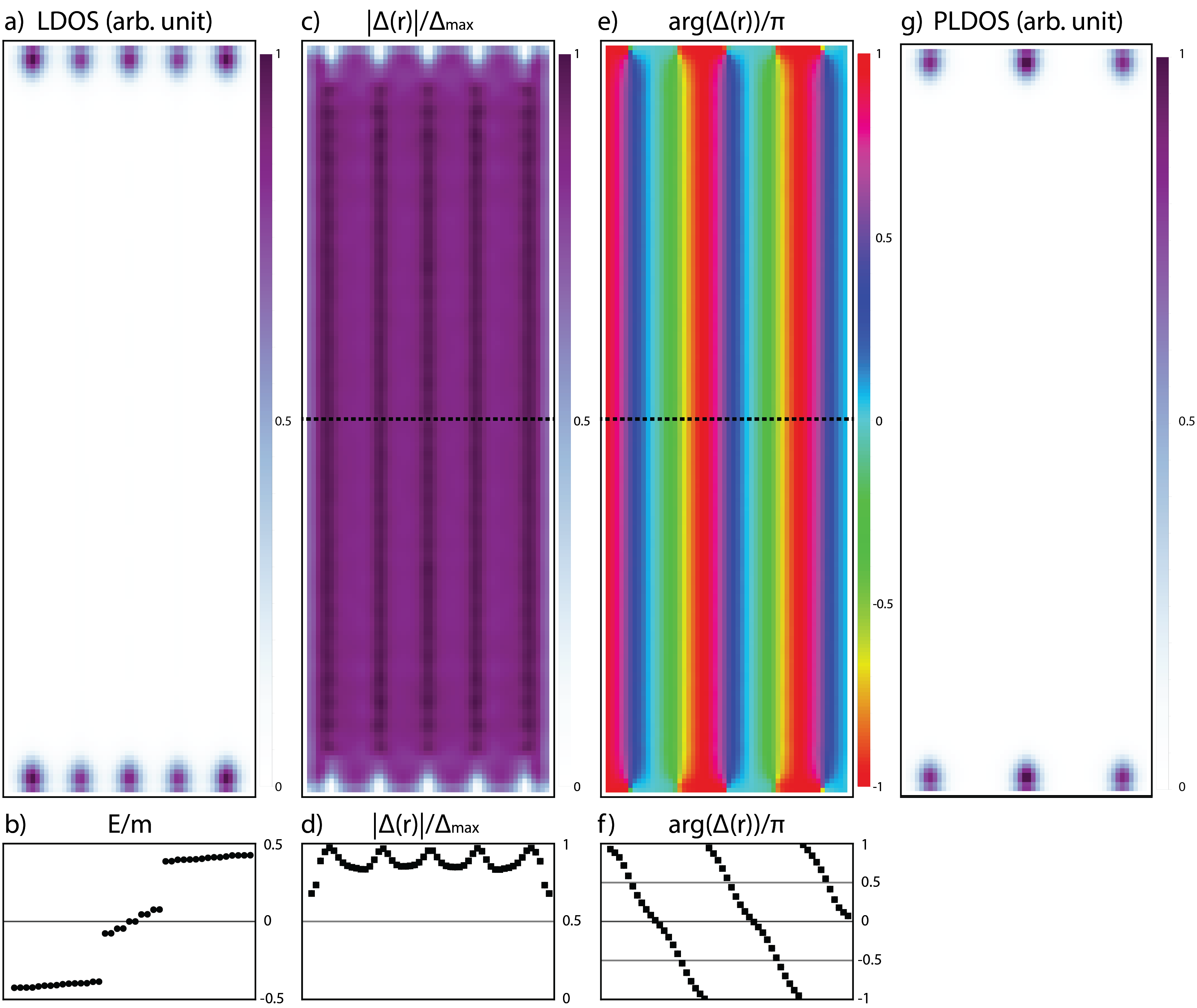} 
\caption{
Topologically nontrivial stripes with order parameter phase $\phi(\mathbf{r})=\pi/2$ with 10 Majorana modes at their ends separated by a distance $\lambda/2=\pi/Q\approx \pi v/2{b}_{xy}$, corresponding to Fig.~2(a) of the main text.
(a) Local density of states (LDOS) at zero energy for a finite system of size $162\times 54$ lattice sites with open boundary conditions.
(b) Energy spectra, showing 10 lowest energy levels below the gap corresponding to the 10 Majorana modes.
Notice the finite energy dispersion due to the finite overlap between Majorana modes.
(c) The magnitude of the order parameter in-plane and (d) along the dotted line.
(e) The argument of the order parameter in-plane and (f) along the dotted line.
(g) Partial LDOS at zero energy of the 2 lowest energy levels.
The order parameter is calculated self-consistently, reaches its maximum $\Delta_\text{max}=0.9m$, and is compatible with $\theta=\pi/4$ in \cref{eq:HFFLO2}.
The topologically nontrivial stripes are realized in correspondence with 1D lines $\phi(\mathbf{r})=\pm\pi/2$.
}
 \label{fig:10majo}
\end{figure}

\begin{figure}[t]
 \centering
 	\includegraphics[height=.6\columnwidth]{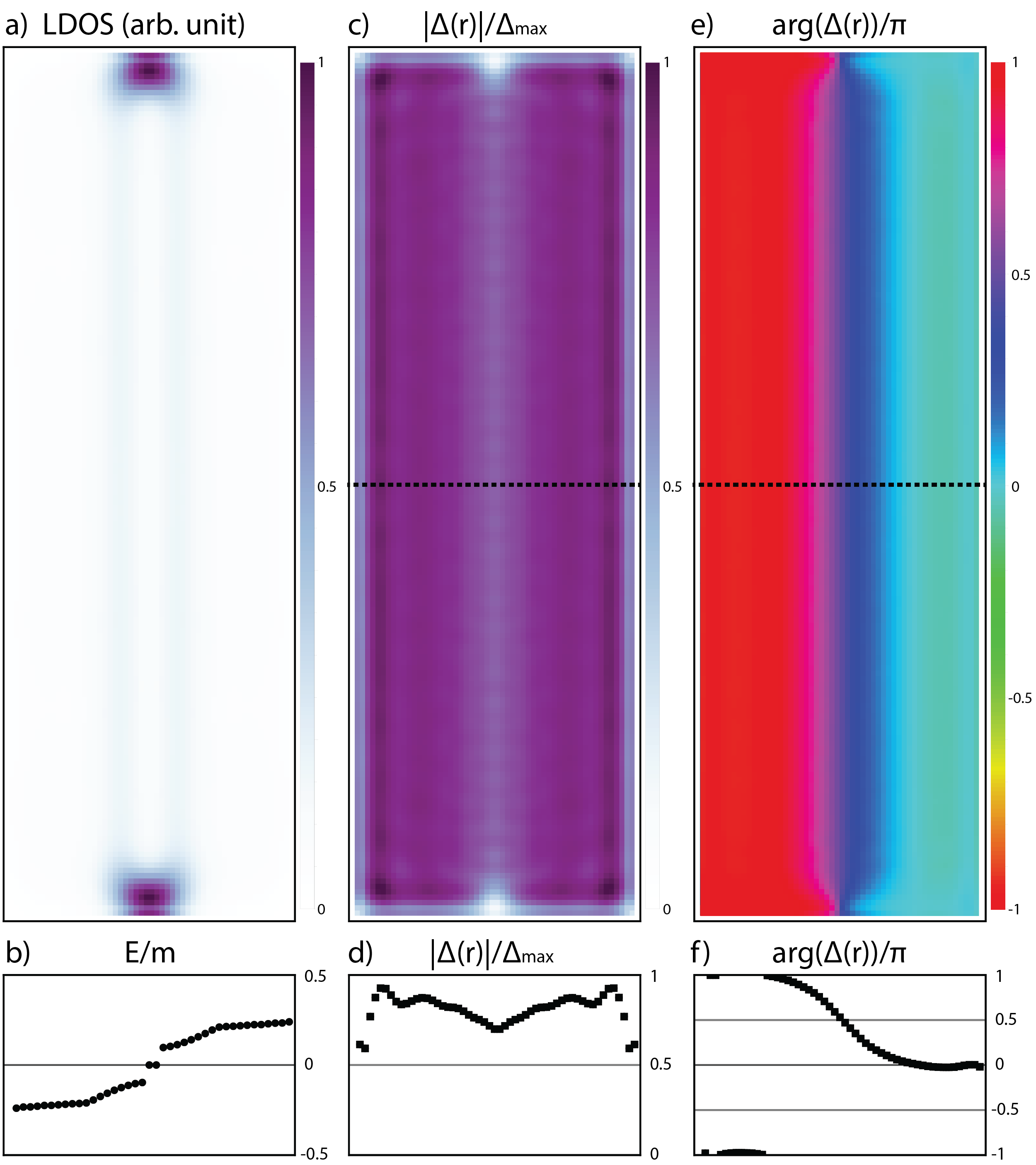} 
\caption{
A single topologically nontrivial stripe with 2 Majorana modes at their ends and corresponding 2 energy levels below the gap, analogously to \cref{fig:10majo}, corresponding to Fig.~2(b) of the main text.
This configuration is obtained by increasing the distance between topologically nontrivial stripes $\lambda/2=\pi/Q\approx \pi v/2{b}_{xy}$ by decreasing the in-plane field component, but keeping the total magnitude of the field unchanged, i.e., by rotating the field direction.
Here, the order parameter reaches the maximum $\Delta_\text{max}=1.2m$.
 }
 \label{fig:2majo}
\end{figure}

\begin{figure}[t]
 \centering
 	\includegraphics[height=.6\columnwidth]{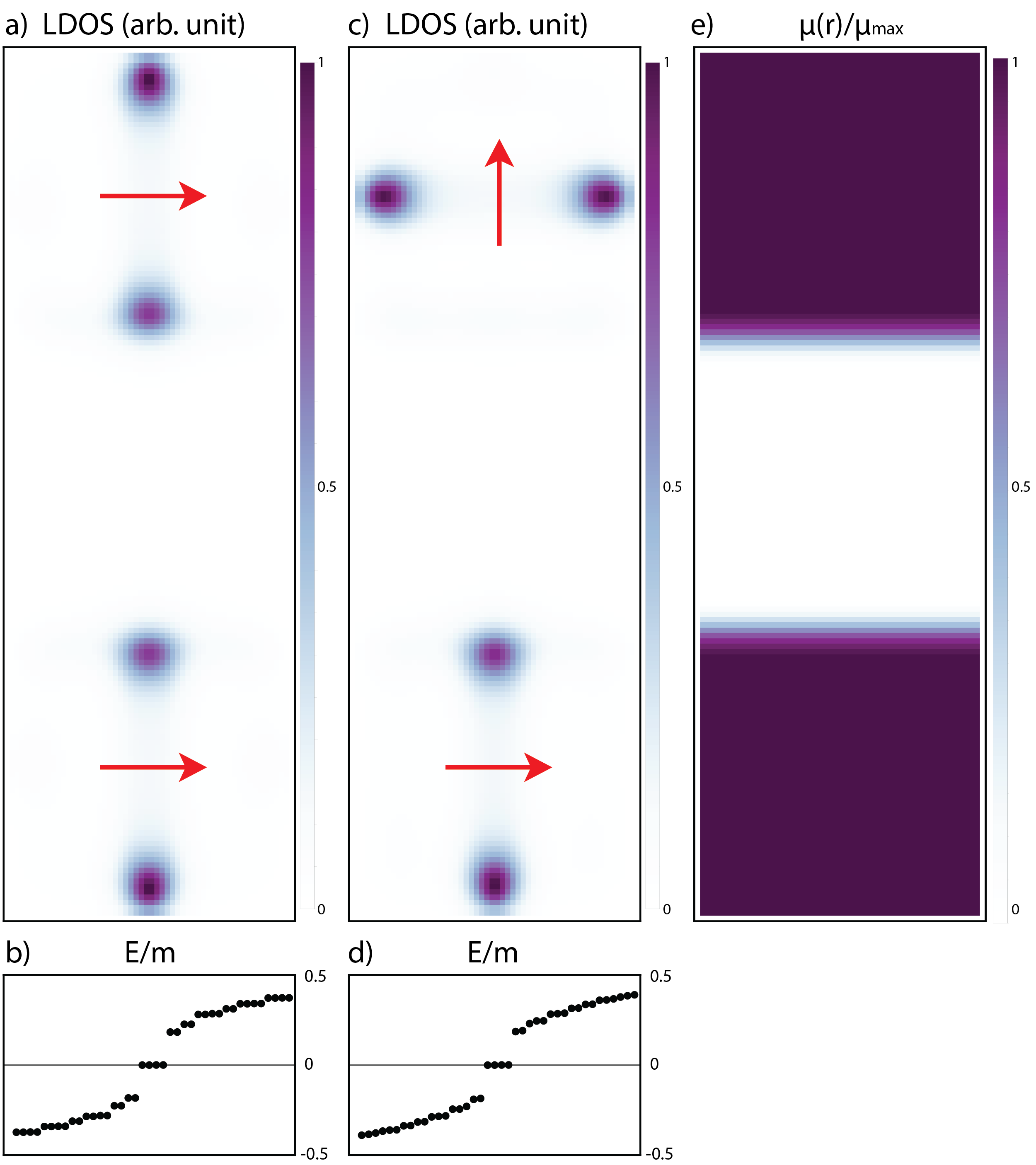} 
\caption{
Two topologically nontrivial stripes with 4 Majorana modes in the presence of a trivial region $\mu=0$ in the middle, corresponding to Figs.~2(c) and (e) of the main text.
(a) Local density of states (LDOS) at zero energy.
(b) The corresponding energy spectra, showing 4 lowest energy levels below the gap.
(c) and (d) same as before, but rotating the in-plane field on the top.
Notice that the energy levels below the gap remain pinned at zero energy.
(e) Spatial dependence of the chemical potential.
Arrows indicate the direction of the 
direction of the Cooper pair momentum $\mathbf{Q}$, which is perpendicular to the
in-plane field $\mathbf{b}_{xy}$. 
}
 \label{fig:4majo}
\end{figure}

\begin{figure}[t]
 \centering
 	\includegraphics[height=.6\columnwidth]{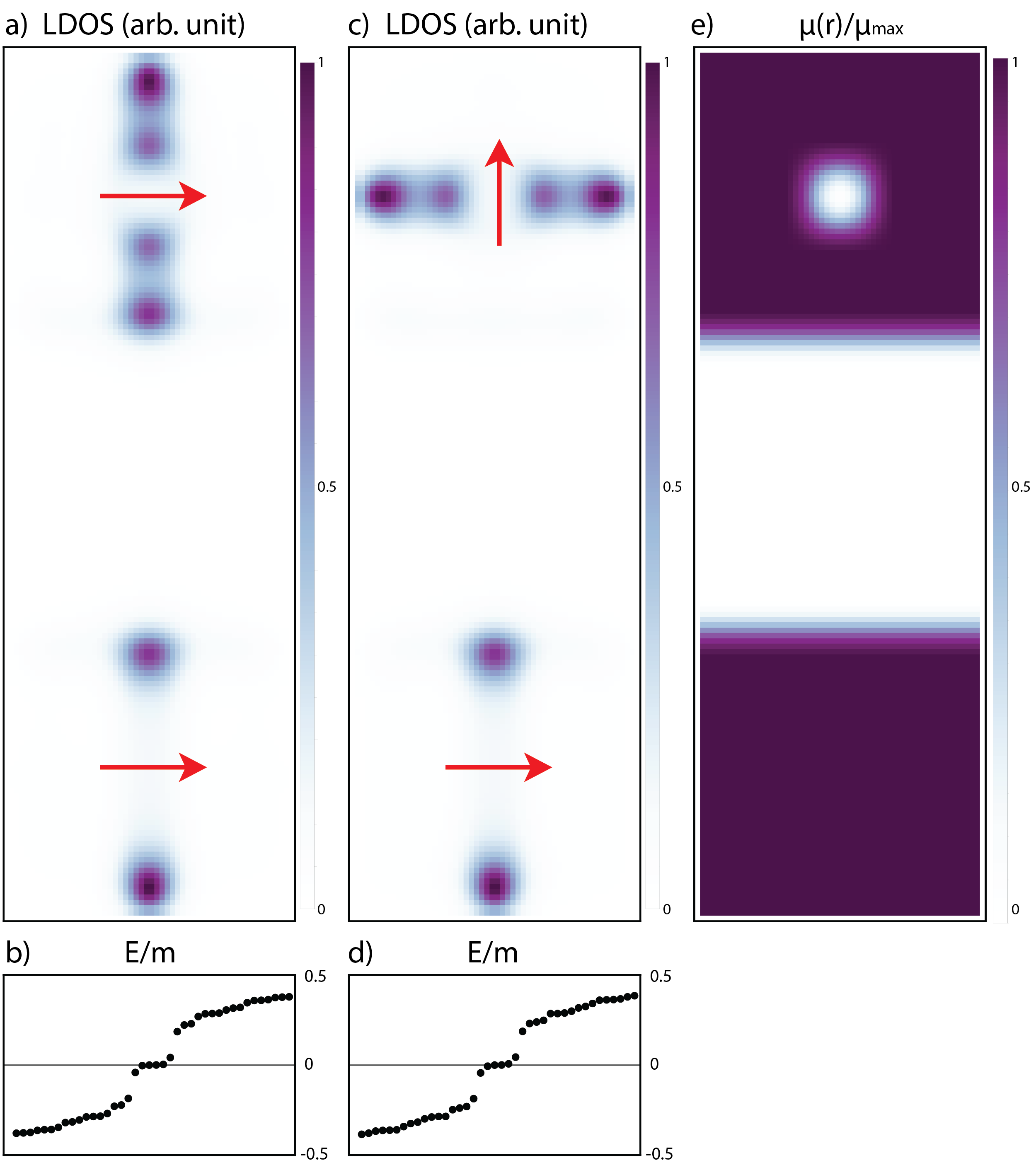} 
\caption{
Three topologically nontrivial stripes with 6 Majorana modes and corresponding 6 energy levels below the gap in the presence of two trivial regions $\mu=0$, analogously to \cref{fig:4majo}, corresponding to Figs.~2(d) and (f) of the main text.
Also in this case the energy levels below the gap remain close to zero energy, with a small energy splitting due to the larger overlap between Majorana modes localized at a closer distance.
}
 \label{fig:6majo}
\end{figure}

We consider now the case of inhomogeneous superconductivity described by the BdG Hamiltonian
\begin{gather}
H_\mathrm{BdG}=
\left[
( m' \mathbf{p}^2 + m)\tau_x
-v\mathbf{p}\times\bm{\sigma}\tau_z
-\mu
\right]
\upsilon_z 
+
\mathbf{b}\cdot\bm{\sigma}
\nonumber\\+
\Re
\left(
\Delta_1(\mathbf{r})\tau_+
+
\Delta_2(\mathbf{r})\tau_-
\right)
\upsilon_x
+
\Im\left(
\Delta_1(\mathbf{r})\tau_+
+
\Delta_2(\mathbf{r})\tau_-
\right)\upsilon_y
,
\end{gather}
where $\tau_\pm=(\tau_x+\ii\tau_y)/2$,
which is equivalent to Eq.~(3) of the main text
if $\Delta_1(\mathbf{r})=\Delta(\mathbf{r})=|\Delta(\mathbf{r})|\ee^{\ii\phi(\mathbf{r})}$ and $\Delta_2(\mathbf{r})=\Delta(\mathbf{r})^*=|\Delta(\mathbf{r})|\ee^{-\ii\phi(\mathbf{r})}$.
The spatially-dependent superconducting order is defined as in Eq.~(5) of the main text
\begin{equation}\label{eq:HFFLO2}
\Delta(\mathbf{r}) = \Delta_0 
\left[
\cos{\theta}\cos{(\mathbf{Q} \cdot \mathbf{r})} + \ii\sin{\theta}\sin{(\mathbf{Q} \cdot \mathbf{r})}
\right],
\end{equation}
where $\Delta_0$ and $\theta$ are determined by the minimum of the free energy, which at zero temperature is $\mathcal F=\langle \mathcal H \rangle$.
A necessary condition for the existence of the minimum is the variational condition $\delta \mathcal F=0$, which allows us to 
find the minima of the free energy self-consistently for any given configuration. 
A simple symmetry argument can help us to find values of $\theta$ that minimize (or maximize) the free energy $\delta \mathcal F=0$.
We first notice that 
\begin{equation}
\Delta(\mathbf{r},\alpha\pm\theta) = \Delta_0 
\left[
(\cos{\alpha}\cos{\theta}\mp\sin{\alpha}\sin{\theta})
\cos{(\mathbf{Q} \cdot \mathbf{r})} 
+\ii
(\sin{\alpha}\cos{\theta}\pm\cos{\alpha}\sin{\theta})
\sin{(\mathbf{Q} \cdot \mathbf{r})}
\right]
,
\end{equation}
which for $\alpha=0,\pi/4$, and $\pi/2$ gives
\begin{subequations}
\begin{equation}
\Delta(\mathbf{r},\pm\theta) = \Delta_0 
\left[
\cos{\theta}
\cos{(\mathbf{Q} \cdot \mathbf{r})} 
\pm\ii
\sin{\theta}
\sin{(\mathbf{Q} \cdot \mathbf{r})}
\right]
,
\end{equation}
\begin{equation}
\Delta(\mathbf{r},\pi/4\pm\theta) = \frac{\Delta_0}{\sqrt2}
\left[
(\cos{\theta}\mp\sin{\theta})
\cos{(\mathbf{Q} \cdot \mathbf{r})} 
+\ii
(\cos{\theta}\pm\sin{\theta})
\sin{(\mathbf{Q} \cdot \mathbf{r})}
\right]
,
\end{equation}
\begin{equation}
\Delta(\mathbf{r},\pi/2\pm\theta) = \Delta_0 
\left[
\mp\sin{\theta}
\cos{(\mathbf{Q} \cdot \mathbf{r})} 
+\ii
\cos{\theta}
\sin{(\mathbf{Q} \cdot \mathbf{r})}
\right]
,
\end{equation}
\end{subequations}
which mandates
\begin{equation}
\Delta(\mathbf{r},-\theta) = \Delta(\mathbf{r},\theta)^* 
,\quad
\Delta(\mathbf{r},\pi/4-\theta) =
\ii
\Delta\left(\frac{\pi\mathbf{Q}}{2Q^2} - \mathbf{r},\pi/4+\theta\right)^*
,\quad
\Delta(\mathbf{r},\pi/2-\theta) =-\Delta(\mathbf{r},\pi/2+\theta)^*
\end{equation}
If $\Delta_1(\mathbf{r})=\Delta(\mathbf{r})$ and $\Delta_2(\mathbf{r})=\Delta(\mathbf{r})^*$, then
taking the complex conjugate of the order parameter correspond to the swapping $\Delta_1(\mathbf{r})\leftrightarrow\Delta_2(\mathbf{r})$, which is equivalent to the unitary transformation $U=\tau_x I$ where $I$ is the spatial inversion $\mathbf{p}\to-\mathbf{p}$, $\mathbf{r}\to-\mathbf{r}$, and $\mathbf{Q}\to-\mathbf{Q}$.
Moreover, multiplying the order parameter by a complex number with of norm one is equivalent to a gauge transformation of the order parameter.
Hence, the equations above mandate that the Hamiltonian $\mathcal H (\theta)$ and $\mathcal H (-\theta)$ are unitary equivalent, as well as the $\mathcal H (\pi/2+\theta)$ and $\mathcal H (\pi/2-\theta)$.
Moreover, the Hamiltonian $\mathcal H (\pi/4+\theta)$ and $\mathcal H (\pi/4-\theta)$ are unitary equivalent, and a combination of translation and inversion $\mathbf{r}\to{\pi\mathbf{Q}}/{(2Q^2)} - \mathbf{r}$, if the boundary conditions are periodic.
Hence,
$ H (\alpha+\theta)$ and $ H (\alpha-\theta)$ are unitarily equivalent, and therefore isospectral,
for $\alpha=0,\pi/4,\pi/2$, and thus
$\mathcal F (\alpha+\theta)=\mathcal F (\alpha-\theta)$
for $\alpha=0,\pi/4,\pi/2$.
Now, for Rolle's theorem, this mandates that there is 
at least one extrema $\delta \mathcal F (\tilde\theta)=0$, 
at least one extrema $\delta \mathcal F (\pi/4+\tilde\theta)=0$, and
at least one extrema $\delta \mathcal F (\pi/2+\tilde\theta)=0$, 
with $\tilde\theta\in=[-\theta,\theta]$.
By taking $\theta\to0$, one obtains that $\delta \mathcal F (0)=0$, $\delta \mathcal F (\pi/4)=0$, and $\delta \mathcal F (\pi/2)=0$.
Hence, $\theta=0,\pi/4,\pi/2$ are local minima (or maxima) of the free energy $\delta \mathcal F=0$.

\subsection{Case $\theta=0$}

The case $\theta=0$ corresponds to a Larkin–Ovchinnikov (LO) order with a constant phase $\phi(\mathbf{r})=0$ and magnitude 
$|\Delta(\mathbf{r})|=\Delta_0|\cos{(\mathbf{Q}\cdot\mathbf{r}})|$,
which becomes zero at the nodal lines $\mathbf{Q}\cdot\mathbf{r}=\pi/2+n\pi$ and reaches its maximum $\Delta_0$ at the antinodal lines $\mathbf{Q}\cdot\mathbf{r}=n\pi$.
Since we have already seen that the energy dispersion is not gapped at finite momenta for $\phi=0$, it is clear that the case $\theta=0$ cannot support a topologically nontrivial gapped phase.

\subsection{Case $\theta=\pi/2$}

The case $\theta=\pi/2$ corresponds to a LO order with a constant phase $\phi(\mathbf{r})=\pi/2$ and magnitude 
$|\Delta(\mathbf{r})|=\Delta_0|\sin{(\mathbf{Q}\cdot\mathbf{r}})|$,
which becomes zero at the nodal lines $\mathbf{Q}\cdot\mathbf{r}=n\pi$ and reaches its maximum $\Delta_0$ at the antinodal lines $\mathbf{Q}\cdot\mathbf{r}=\pi/2+n\pi$.
The topological invariant
on the nodal and antinodal lines, respectively, becomes 
\begin{subequations}
\begin{align}\label{eq:TInodal}
(-1)^{\nu}&=\sgn
\prod_{\pm\pm}
(b\pm m\pm\mu),
\\\label{eq:TIantinodal}
(-1)^{\nu}&=\sgn
\prod_{\pm\pm}
\big(b\pm m\pm\sqrt{\Delta_0^2+\mu^2}\big).
\end{align}
\end{subequations}
Topologically nontrivial stripes are realized when the topological invariant assumes different values on the nodal and antinodal lines.
Only antinodal lines can realize a 1D topological superconductor since nodal lines are not superconducting ($\Delta=0$ on nodal lines).
We thus require that $\nu=0$ on nodal lines and $\nu=1$ on antinodal lines.
\Cref{eq:TInodal} gives $\nu=0$ on nodal lines for $b>|m|+|\mu|$, $|\mu|>b+|m|$, or $|m|>b+|\mu|$.
On the other hand, \cref{eq:TIantinodal} gives $\nu=1$ on antinodal lines for $|m|+b>\sqrt{\Delta^2+\mu^2}>| |m|-b |$.
These two conditions leave only one possibility, i.e., $|m|+b>\sqrt{\Delta^2+\mu^2}>| |m|-b |>|\mu|$
However, this is not compatible with the presence of a superconducting state, which requires the Fermi level to lie within the conduction band, which is only possible for $|\mu|>b-|m|$.
Hence, the case $\theta=\pi/2$ cannot realize topologically nontrivial stripes.

\subsection{Case $\theta=\pi/4$}

The case $\theta=\pi/4$ corresponds to a Fulde-Ferrel (FF) order with a constant magnitude $\Delta_0/\sqrt2$ and a phase $\phi(\mathbf{r})=\mathbf{Q}\cdot\mathbf{r}$ giving $\sin^2(\phi(\mathbf{r}))=0,1$
respectively for $\mathbf{Q}\cdot\mathbf{r}=n\pi$ and $\mathbf{Q}\cdot\mathbf{r}=\pi/2+n\pi$.
The topological invariant for $\mathbf{Q}\cdot\mathbf{r}=n\pi,\pi/2+n\pi$, respectively, becomes
\begin{subequations}
\begin{align}
(-1)^{\nu}&=
\left|m^2+\mu^2+\Delta_0^2/2-b^2\right|-2|\mu|| m|,
\\
(-1)^{\nu}&=
\left|m^2+\mu^2+\Delta_0^2/2-b^2\right|-2|m|\sqrt{\Delta_0^2/2+\mu^2}.
\end{align}
\end{subequations}
Topologically nontrivial stripes are realized when the topological invariant assumes different values on $\mathbf{Q}\cdot\mathbf{r}=n\pi,\pi/2+n\pi$.
This is only possible when $\nu=0$ at $\mathbf{Q}\cdot\mathbf{r}=n\pi$ and $\nu=1$ at $\mathbf{Q}\cdot\mathbf{r}=n\pi,\pi/2+n\pi$, if
\begin{equation}
2|m|\sqrt{\Delta_0^2/2+\mu^2}>\left|m^2+\mu^2+\Delta_0^2/2-b^2\right|>2|\mu|| m|,
\end{equation}
which is Eq.~(6) of the main text.

\section*{Supplementary Note 4. Numerical calculations with lattice Hamiltonians}

To verify the presence of the gapless hinge modes in the different regimes, we discretize the Hamiltonian on a discrete 2D lattice and calculate numerically the LDOS at zero energy for a finite system.
The effective Hamiltonian describing the surface states can be regularized in real space by discretization on a 2D square lattice by taking $\mathbf{r}_{n,m}=n \mathbf{a}_x + m \mathbf{a}_y$ with $\mathbf{a}_x,\mathbf{a}_y$ the lattice vectors of the lattice, approximating the integral with discrete sums, and defining the differential operators in terms of finite differences, taking
\begin{subequations}
\begin{align}
p_x		&\to	-\frac{\ii\hbar}{2a} \left(c_{n+1,m}-c_{n-1,m}\right)c_{nm},
\\
p_x^2	&\to	-\frac{\hbar^2}{a^2} \left(c_{n+1,m}-2c_{n,m}+c_{n-1,m}\right)c_{nm},
\\
p_y		&\to	-\frac{\ii\hbar}{2a} \left(c_{n,m+1}-c_{n,m-1}\right)c_{nm},
\\
p_y^2	&\to	-\frac{\hbar^2}{a^2} \left(c_{n,m+1}-2c_{n,m}+c_{n,m-1}\right)c_{nm},
\end{align}
\end{subequations}
where $c_{nm}=\psi^\dag(\mathbf{r}_{nm})$.

We diagonalize numerically the Hamiltonian, obtaining the energy levels $E_\eta$ and eigenstates in the form
\begin{equation}
\psi_\eta=\sum_{\mathbf{r}\tau\sigma} u_{\eta\tau\sigma}(\mathbf{r})\psi_{\tau\sigma}(\mathbf{r}) + v_{\eta\tau\sigma}(\mathbf{r})\psi^\dag_{\tau\sigma}(\mathbf{r}),
\end{equation}
and consequently calculate LDOS at zero energy as
\begin{equation}
\rho(\mathbf{r})=\sum_{\eta\tau\sigma} 
\left(|u_{\eta\tau\sigma}(\mathbf{r})|^2 + |v_{\eta\tau\sigma}(\mathbf{r})|^2\right)
\frac1\pi\Im\left(\frac1{E_\eta+\ii\Gamma}\right),
\end{equation}
with $\Gamma$ a smearing parameter.

In order to perform the self-consistent numerical calculations, we consider a BdG Hamiltonian with independent superconducting order parameters $\Delta_{\tau}(\mathbf{r})$ on the top and bottom surface states $\tau=1,2$, given by
\begin{gather}
H_\mathrm{BdG}(\mathbf{r})=
\left[v(p_y \sigma_x - p_x \sigma_y)\tau_z + 
( m' p^2 + m)\tau_x
\right]
\upsilon_z 
+
\mathbf{b}\cdot\bm{\sigma}
+
\begin{bmatrix}
0&0&\Delta_1(\mathbf{r})&0\\
0&0&0&\Delta_2(\mathbf{r})\\
\Delta_1(\mathbf{r})^*&0&0&0\\
&\Delta_2(\mathbf{r})^*&0&0\\
\end{bmatrix}
,
\end{gather}
where 
${\psi}(\mathbf{r})=\left[{\psi}_{1,\up}(\mathbf{r}),{\psi}_{1,\down}(\mathbf{r}),{\psi}_{2,\up}(\mathbf{r}),{\psi}_{2,\down}(\mathbf{r})\right]^\intercal$.

To determine the energy spectra and the superconducting order parameter self-consistently, we first make the ansatz $\Delta_{1,2}(\mathbf{r})=\Re(\Delta(\mathbf{r}))\pm\ii\,\Im(\Delta(\mathbf{r}))$ where $\Delta(\mathbf{r})$ is given by \cref{eq:HFFLO2} (i.e., Eq.~(5) of the main text) with $Q=2b_{xy}/v$
and $\mathbf{Q}\perp \mathbf{b}_{xy}$,
$\Delta_0=m$, and several choices of $\theta=0,\pi/8,\pi/4,3\pi/8,\pi/2$.
We then diagonalize numerically the discrete Hamiltonian and obtain the energy eigenvalues $E_\eta$ and the eigenstates in the form
\begin{equation}
\psi_\eta=\sum_{\mathbf{r}\tau\sigma} u_{\eta\tau\sigma}(\mathbf{r})\psi_{\tau\sigma}(\mathbf{r}) + v_{\eta\tau\sigma}(\mathbf{r})\psi^\dag_{\tau\sigma}(\mathbf{r}),
\end{equation}
where the sum is on the discrete lattice $\mathbf{r}=\mathbf{r}_{nm}$.

We then calculate the superconducting order parameter at zero temperature, separately for the top and bottom surface states 
and as a function of the spatial coordinate 
using
\begin{gather}\label{eq:selfconsistent}
\Delta_{\tau}(\mathbf{r})
=
-U
\bra{\text{0}}
{\psi}_{\tau\down}(\mathbf{r})
{\psi}_{\tau\up}(\mathbf{r})
\ket{\text{0}}
=
-U\sum_\eta u^*_{\eta\tau\down}(\mathbf{r}) v_{\eta\tau\up}(\mathbf{r})
,
\end{gather}
where 
$U$ is the effective electron-electron coupling constant.
We then close the self-consistency loop by plugging the obtained order parameter above into the Hamiltonian, diagonalize, and calculate again the superconducting order parameter $\Delta_{\tau}(\mathbf{r})$ 
and as a function of the spatial coordinate 
using \cref{eq:selfconsistent}, iterating the process to obtain successive approximations $(\Delta_{\tau}(\mathbf{r}))_1,(\Delta_{\tau}(\mathbf{r}))_2,\ldots$, until the order parameter converges, 
such that the standard deviation satisfies
\begin{equation}
\frac1{\sqrt{N}} \| (\Delta_{\tau}(\mathbf{r}))_n-(\Delta_{\tau}(\mathbf{r}))_{n-1} \|
=\left[\frac1N 
\sum_\mathbf{r} |(\Delta_{\tau}(\mathbf{r}))_n-(\Delta_{\tau}(\mathbf{r}))_{n-1}|^2
\right]^{1/2}
<0.05 m,
\end{equation}
where $N$ is the total number of lattice points. 
Since we run calculations for several initial choices of $\theta=0,\pi/8,\pi/4,3\pi/8,\pi/2$, we then select, among the calculations which converge, the order parameter $\Delta_{\tau}(\mathbf{r})$ corresponding to the smallest free energy at zero temperature, which is
\begin{equation}
\mathcal F=\langle \mathcal H\rangle=\sum_{E_\eta<0} E_\eta + \frac1U\sum_\mathbf{r}\sum_\tau |\Delta_{\tau}(\mathbf{r})|^2.
\end{equation}
After that, we determine the value of $\theta$ in \cref{eq:HFFLO2} (i.e., Eq.~(5) of the main text) which best fit the order parameter (which does not necessarily coincide with the initial choice).
For all instances, we found that $\theta=\pi/4$ is the best fit.

We use $m'=m$,
Dirac velocity $v=3.3m$,
magnetic field magnitude $b=0.5m$, 
and chemical potential $\mu=m$, excluding the regions where $\mu=0$.
(effective lattice parameter $a=\text{\SI{50}{\nano\meter}}$).
In all the chosen configurations of Fig.~2 of the main text, we obtain an order parameter which is approximately equal to \cref{eq:HFFLO2} (i.e., Eq.~(5) of the main text) with $Q\approx2b_{xy}/v$
and $\mathbf{Q}\perp \mathbf{b}_{xy}$,
$\theta=\pi/4$, and magnitude of the order parameter reaching its maximum 
$\Delta_\text{max}=\max|\Delta(\mathbf{r})|\approx1.2m$, 
taking $U=300m$. 
These values correspond to 
$v=\text{\SI{2.67}{\electronvolt\angstrom}}$,
$m=\text{\SI{1.5}{\milli\electronvolt}}$, 
$m'=\text{\SI{500}{\electronvolt\angstrom^2}}$,
$b=\text{\SI{0.9}{\milli\electronvolt}}$ ($B=\text{\SI{2}{\tesla}}$ with $g=15$),
$\mu=\text{\SI{1.5}{\milli\electronvolt}}$ (excluding the regions where $\mu=0$) and
$\Delta_0=\text{\SI{2}{\milli\electronvolt}}$ 
and $U=\text{\SI{0.45}{\electronvolt}}$.
These parameters are compatible with heterostructures of
\ce{Bi2Te3}\cite{chen_experimental_2009} proximitized with \ce{NbTiN} or \ce{NbSe2}.

The energy spectra and the spatial dependence of the order parameter corresponding to Fig.~2 of the main text are shown in \cref{fig:10majo,fig:2majo,fig:4majo,fig:6majo}.
The resulting self-consistent order parameter at zero temperature is approximately equal to \cref{eq:HFFLO2} with $\theta=\pi/4$, $Q\approx2b_{xy}/v$, and $\mathbf{Q}\perp \mathbf{b}_{xy}$, corresponding to an FF order with almost constant magnitude and nonuniform phase.
However, there are slight variations from \cref{eq:HFFLO2}, in particular in the regions close to the boundaries, where the magnitude of the order parameter is slightly suppressed.

\section*{Supplementary Note 5. Alternative braiding protocols}

\begin{figure}[t]
\centering
\includegraphics[width=\columnwidth]{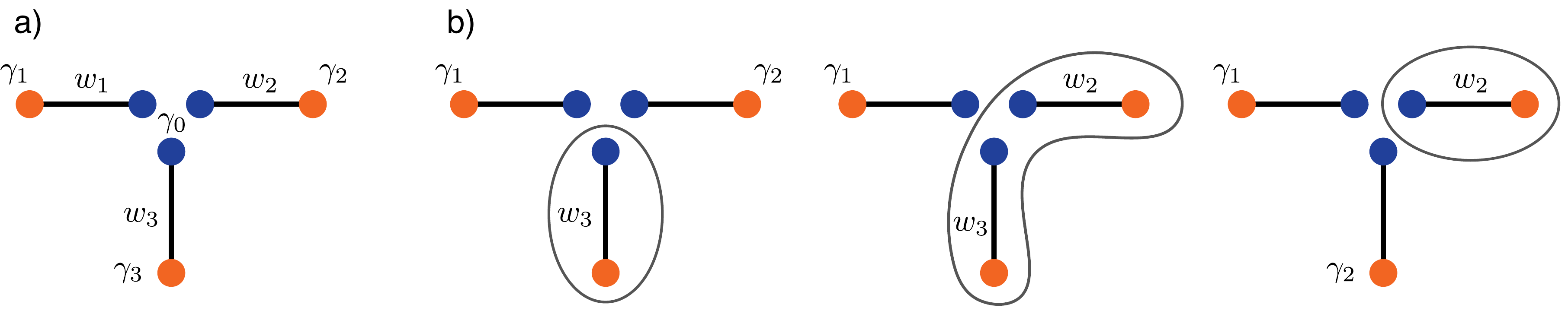} 
\caption{
Braiding Majorana modes in parameter space without moving them.
(a) Three topologically nontrivial stripes arranged around a pointlike defect.
(b)
Moving the Majorana mode $\gamma_2$ from the stripe on the right to the stripe on the bottom (left to right), by controlling the coupling between Majorana modes on opposite ends of the stripes.
The exchange of Majorana modes is obtained by moving them clockwise in a three-step process.
}
\label{fig:braid}
\end{figure}

An alternative way to braid Majorana modes at the end of topologically nontrivial stripes is to exchange them in parameter space, without physically moving the topologically nontrivial stripes, controlling the coupling between Majorana modes\cite{sau_controlling_2011,karzig_shortcuts_2015}.
Let us consider three TNTs arranged around a pointlike defect as in \cref{fig:braid}(a).
The hybridization of the three Majorana modes at the inner ends of the three TNTs results into an unpaired Majorana mode at the center\cite{sau_controlling_2011,karzig_shortcuts_2015}.
Hence, there are three Majorana zero modes $\gamma_1$, $\gamma_2$, and $\gamma_3$ localized on the outer ends of the TNTs, and another one $\gamma_0$ at the center.
This setup is analogous to a trijunction, described in Refs.~\cite{sau_controlling_2011,karzig_shortcuts_2015}, and corresponds to the effective Hamiltonian
\begin{equation}
	H=\ii\sum_{n=1}^3 w_n\gamma_0\gamma_n,
\end{equation}
where $w_n\propto\ee^{-L_{n}/\xi_n}$ are the couplings between the central mode $\gamma_0$ and the remaining Majorana modes, with $L_n$ and $\xi_n$ respectively the length of the TNTs and the Majorana localization length on each TNTs.
These couplings can be controlled by increasing or decreasing the length of the TNTs or changing the Majorana localization lengths on each TNNs by locally varying the the magnetic field or other parameters.
To braid the Majorana modes, we simply apply the braiding protocols described in Refs.~\cite{sau_controlling_2011,karzig_shortcuts_2015} to our case.
This protocol is sketched in \cref{fig:braid}(b).
We start with a configuration with $w_1,w_2\approx0$ and $w_3>0$, where $\gamma_1$ and $\gamma_2$ are fully decoupled and have zero energy, while the modes $\gamma_0$ and $\gamma_3$ are hybridized at finite energy.
Next, we increase the coupling $w_2$ to a finite value, leaving the mode $\gamma_1$ unaffected and hybridizing the modes $\gamma_1$, $\gamma_2$, and $\gamma_3$.
Then, we reduce the coupling $w_3$, such that $w_1,w_3\approx0$ and $w_2>0$.
By doing this, the mode $\gamma_2$ localizes on the topologically nontrivial stripe on the bottom.
This process amounts to moving the Majorana mode $\gamma_2$ from the stripe on the right to the stripe on the bottom.
The exchange of the two Majorana modes $\gamma_1$ and $\gamma_2$ is then realized by three consecutive processes, i.e., by moving the mode $\gamma_2$ from the stripe on the right to the stripe on the bottom, then $\gamma_1$ from the stripe on the left to the stripe on the right, and finally, $\gamma_2$ from the stripe on the bottom to the stripe on the left.
The whole process amounts to exchanging the two Majorana modes $\gamma_1$ and $\gamma_2$.
This braiding process is realized without moving the topologically nontrivial stripes, and only requires the ability to change the coupling between Majorana modes at the end of the stripes.
This can be achieved by changing the length of the stripes, or varying the effective Majorana localization length on each stripe, which can be controlled by varying the out-of-plane magnetic field or other parameters.

\section*{Supplementary Note 6. Zero energy modes in a set of $N$ topologically nontrivial stripes}

A set of $2N$ Majorana modes localized at the ends of $N$ topologically nontrivial stripes can be effectively described by the Hamiltonian
\begin{equation}
H_{N}=
\ii w\sum_{n=1}^{N-1}
\left(
\gamma_{L,n} \gamma_{L,n+1}
+
\gamma_{R,n} \gamma_{R,n+1}
\right)
+\ii w' \sum_{n=1}^N \gamma_{L,n} \gamma_{R,n}.
\end{equation}
where $\gamma_{L,n}$ and $\gamma_{R,n}$ with $n=1,\ldots,N$ are the modes on the left and right ends of the stripes, respectively, $w$ the couplings between modes on the same side, and $w'$ the coupling between modes on opposite sides.

\subsection{Case $w'=0$ with $N=2m+1\ge1$}

If $w'=0$ and the number of modes is odd $N=2m+1\ge1$, the energy spectrum of the Hamiltonian $H_{N}$ has two zero energy modes $\widetilde\gamma_1$ and $\widetilde\gamma_2$, linear combinations of the Majorana modes $\gamma_n$.
In this case, 
the $\gamma_{L,n}$ and $\gamma_{R,n}$ operators decouple and thus the Hamiltonian can be written in matrix form as a block matrix given by
\begin{equation}
H_{N}=
\ii w\sum_{n=1}^{N-1}
\left(
\gamma_{L,n} \gamma_{L,n+1}
+
\gamma_{R,n} \gamma_{R,n+1}
\right)
=
\frac{\ii w}2 
\begin{bmatrix}
 B_{N} & 0 \\
 0 & B_{N} \\
\end{bmatrix},
\end{equation}
with the antisymmetric matrix $B_{N}$ of order $N$ given by
\begin{equation}
B_{N}=
\begin{bmatrix}
 0 & 1 & \\
 -1 & 0 & 1 \\
 & -1 & 0 & \ddots \\
 & & \ddots & \ddots & 1\\
 & & & -1 & 0 \\
\end{bmatrix},
\end{equation}
which is a tridiagonal matrix with zero on the main diagonal.
One has that $\det B_{N+2}=\det B_{N}$ for the properties of the determinant of tridiagonal matrices.
If the number of modes is even $N=2m\ge2$, then 
$\det B_{2}=1$, 
which mandates by induction $\det B_{2m}>0$.
Therefore, the Hamiltonian $H_{N}$ with even $N=2m\ge2$ does not have zero eigenvalues.
If the number of modes is odd $N=2m+1\ge1$, then $\det B_{1}=0$, which mandates by induction $\det B_{2m+1}=0$ for all $m\ge1$.
Therefore, the Hamiltonian $H_{N}$ with odd $N=2m+1\ge1$ has at least two zero eigenvalues, one for each block.
Moreover, due to the particle-hole symmetry of the Hamiltonians on each block, the number of positive energy levels $E>0$ is equal to the number of negative energy levels $E<0$.
Since the total number of eigenvalues for each block $N=2m+1$ is odd, the number of zero energy modes for each block must also be odd.
Moreover, direct numerical calculations indicate that there are only two zero energy modes in total. 
Hence, we conclude that the Hamiltonian $H_{N}$ with odd $N=2m+1\ge1$ has two and only two zero eigenvalues, one for each block.
These two eigenvalues correspond to the normalized eigenstates 
\begin{equation}
\widetilde\gamma_{1}={\frac{1}{\sqrt{m+1}}}\sum_{n=0}^{m} \gamma_{L,2n+1},
\qquad
\widetilde\gamma_{2}={\frac{1}{\sqrt{m+1}}}\sum_{n=0}^{m} \gamma_{R,2n+1},
\end{equation}
as one can verify by direct substitution.
The case $w'=0$ is approximately realized when the distance between the Majorana modes at the left and right edges of the system is larger than the distance between contiguous Majorana modes on the same side, as is the case in \cref{fig:10majo} and Fig.~2(a) of the main text.
The partial LDOS at zero energy of the lowest energy levels $\pm E\approx0$ is shown in \cref{fig:10majo}(g).

\subsection{Case $|w|=|w'|$ with $N=3m+2\ge2$}

If $w=w'$ and the number of modes is $N=3m+2\ge2$, the energy spectrum of the Hamiltonian $H_{N}$ has two zero energy modes $\widetilde\gamma_1$ and $\widetilde\gamma_2$, linear combinations of the modes $\gamma_n$.
It is possible to prove this by induction.
For $N=2$, one has
\begin{equation}\label{eq:H2}
H_{2}=	
 \ii w \left( \gamma_{L1} \gamma_{L2} + \gamma_{R1} \gamma_{R2} 
+ \gamma_{L1} \gamma_{R1} + \gamma_{L2} \gamma_{R2} \right)
=	 \frac{\ii w}2 A_2=
\frac{\ii w}2 
\begin{bmatrix}
 0 & 1 & 1 & 0 \\
-1 & 0 & 0 & 1 \\
-1 & 0 & 0 & 1 \\
 0 & -1 & -1 & 0 \\
\end{bmatrix},
\end{equation}
where the matrix $A_2$ is antisymmetric.
The Hamiltonian $H_2$ has two zero eigenvalues corresponding to the eigenstates 
\begin{equation}\label{eq:modesH2}
\widetilde\gamma_1= \frac{1}{\sqrt2}\left(\gamma_{L1} + \gamma_{R2}\right),
\qquad
\widetilde\gamma_2= \frac{1}{\sqrt2}\left(\gamma_{R1} - \gamma_{L2}\right),
\end{equation}
as one can verify by direct substitution.

\begin{figure}[t]
\includegraphics[width=\columnwidth]{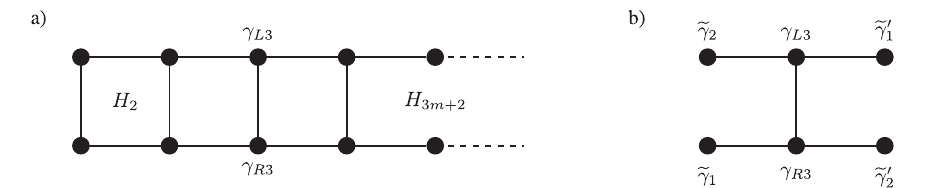} 
\centering
\caption{
Diagrammatic sketch of the Hamiltonian $H_{3(m+1)+2}$.
(a) The Hamiltonian in \cref{eq:H31st} can be 
written in terms of the sum of the terms
$H_2$ 
describing the couplings within the first 4 modes $\gamma_{L,n}$ and $\gamma_{R,n}$ with $n=1,2$, 
$H_{3m+2}$ 
describing the couplings within the last $2(3m+2)$ modes $\gamma_{L,n}$ and $\gamma_{R,n}$ with $n=4,\ldots,3(m+1)+2$,
and additional terms describing the coupling between the 2 modes $\gamma_{L3}$ and $\gamma_{R3}$ and the rest of the system.
(b) The same Hamiltonian can be rewritten as \cref{eq:H3}, which
describes the couplings between
the modes $\widetilde\gamma_1$ and $\widetilde\gamma_2$
(which are the zero modes of $H_2$),
the modes $\widetilde\gamma_1'$ and $\widetilde\gamma_2'$
(which are the zero modes of $H_{3m+2}$),
and the modes $\gamma_{L3}$ and $\gamma_{R3}$.
}
\label{fig:susy}
\end{figure}

To complete the proof by induction, we assume that $H_{3m+2}$ for $m\ge0$ has at least two energy zero modes in the form 
\begin{subequations}\label{eq:modesinduction}
\begin{align}
\widetilde\gamma_1''=& \frac1{\sqrt{2m+2}}\sum_{n=0}^m
\kappa_{n}
\left(\gamma_{L,3n+1} + \gamma_{R,3n+2}\right)
+
\kappa_{n-1}
\left(\gamma_{R,3n+1} - \gamma_{L,3n+2}\right),
\\
\widetilde\gamma_2''=& \frac1{\sqrt{2m+2}}\sum_{n=0}^m
\kappa_{n}
\left(\gamma_{R,3n+1} - \gamma_{L,3n+2}\right)
+
\kappa_{n+1}
\left(\gamma_{L,3n+1} + \gamma_{R,3n+2}\right)
,
\end{align}
\end{subequations}
with $\kappa_n=\cos{\left({n\pi}/2\right)}=1,0,-1,0,\ldots$, and demonstrate that $H_{3(m+1)+2}$ has the same property.
The Hamiltonian $H_{3(m+1)+2}=(\ii w/2) A_{3(m+1)+2}$ can be written in terms of three contributions:
a term $H_{2}=(\ii w/2) A_{2}$ describing the couplings within the first 4 modes $\gamma_{L,n}$ and $\gamma_{R,n}$ with $n=1,2$,
a term $H_{3m+2}=(\ii w/2) A_{3m+2}$ describing the couplings within the last $2(3m+2)$ modes $\gamma_{L,n}$ and $\gamma_{R,n}$ with $n=4,\ldots,3(m+1)+2$,
and additional terms $(\ii w/2)W_{2}$ and $(\ii w/2)W_{3m+2}$ describing the coupling between the 2 modes $\gamma_{L3}$ and $\gamma_{R3}$ and the rest of the system.
In matrix form 
\begin{equation}\label{eq:H31st}
H_{3(m+1)+2}=	 \frac\ii2 w 
\begin{bmatrix}
 A_{3m+2} & \multicolumn{2}{c}{W_{3m+2}} & 0_{3m+2,2} \\
 \multirow{2}{*}{$-W_{3m+2}^\dag$} & 0 & +1 & \multirow{2}{*}{$W_{2}$} \\
 & -1 & 0 & \\
 0_{2,3m+2} & \multicolumn{2}{c}{-W_{2}^\dag} & A_{2} \\
\end{bmatrix},
\end{equation}
where $0_{jk}$ is the 
$2j\times 2k$ zero matrix.
This matrix is represented diagrammatically in \cref{fig:susy}(a).
The spectrum of $H_2$ has two zero-energy modes $\widetilde\gamma_1$ and $\widetilde\gamma_2$ as in \cref{eq:modesH2}
and $H_{3m+2}$ has two zero-energy modes as in \cref{eq:modesinduction} but translated by 3 lattice sites, i.e.,
\begin{subequations}\label{eq:modesinduction2}
\begin{align}
\widetilde\gamma_1'=& \frac1{\sqrt{2m+2}}\sum_{n=1}^{m+1}
\kappa_{n-1}
\left(\gamma_{L,3n+1} + \gamma_{R,3n+2}\right)
+
\kappa_{n-2}
\left(\gamma_{R,3n+1} - \gamma_{L,3n+2}\right),
\\
\widetilde\gamma_2'=& \frac1{\sqrt{2m+2}}\sum_{n=1}^{m+1}
\kappa_{n-1}
\left(\gamma_{R,3n+1} - \gamma_{L,3n+2}\right)
+
\kappa_{n}
\left(\gamma_{L,3n+1} + \gamma_{R,3n+2}\right)
,
\end{align}
\end{subequations}
We do not necessary to determine all the matrix elements in $W_{2}$ and $W_{3m+2}$, but it is sufficient to calculate the matrix elements between the zero energy modes $\widetilde\gamma_{1,2},\widetilde\gamma_{1,2}'$ and $\gamma_{L3},\gamma_{R3}$.
These matrix elements are easily obtained:
The mode $\widetilde\gamma_1$ is coupled to $\gamma_{R3}$ via $\gamma_{R2}$ with matrix element $\ii w/{\sqrt2}$,
the mode $\widetilde\gamma_2$ is coupled to $\gamma_{L3}$ via $\gamma_{L2}$ with matrix element $-\ii w/{\sqrt2}$, 
the mode $\widetilde\gamma_1'$ is coupled to $\gamma_{L3}$ via $\gamma_{L4}$ with matrix element $-\ii w/{\sqrt{2m+2}}$, and
the mode $\widetilde\gamma_2'$ is coupled to $\gamma_{R3}$ via $\gamma_{R4}$ with matrix element $-\ii w/{\sqrt{2m+2}}$. 

Reordering the eigenstates such that the first 6 diagonal elements are the zero eigenvalues corresponding to the eigenstates $\widetilde\gamma_1$, $\widetilde\gamma_2$, $\widetilde\gamma_1'$, $\widetilde\gamma_2'$, followed by the eigenstates 
$\gamma_{L3}$, $\gamma_{R3}$ 
in this order, which yields
\begin{equation}\label{eq:H3}
H_{3(m+1)+2}= \frac\ii2 w 
\begin{bmatrix}
 0 &0 &0 &0 & 0 & \frac{1}{\sqrt2} & \\
 0 &0 &0 &0 & -\frac{1}{\sqrt2} & 0 & \multirow{2}{*}{$0_{2,3m+2}$} \\
 0 &0 &0 &0 & -\frac{1}{\sqrt{2m+2}} & 0 & \\
 0 &0 &0 &0 & 0 & -\frac{1}{\sqrt{2m+2}} & \\
 0 & \frac{1}{\sqrt2} &\frac{1}{\sqrt{2m+2}} &0 & 0 & +1 & \multirow{2}{*}{$W'$} \\
 -\frac{1}{\sqrt2} &0 &0 &\frac{1}{\sqrt{2m+2}} & -1 & 0 & \\
 & \multicolumn{2}{c}{0_{3m+2,4}} & & \multicolumn{2}{c}{-W^{\prime\dag}} & H' \\
\end{bmatrix},
\end{equation}
where $H'$ is the diagonal matrix corresponding to the remaining $6m+4$ higher-energy eigenvalues of the matrices $A_{3m+2}$ and $A_{2}$ on its diagonal, and $W'$ the remaining couplings between the modes 
$\gamma_{L3}$, $\gamma_{R3}$ 
and the rest of the system.
This matrix is represented diagrammatically in \cref{fig:susy}(b).
By using the Laplace expansion on the first row, it is easy to prove that the determinant of the matrix $H_{3(m+1)+2}$ is zero.
Therefore, there is at least one zero eigenvalue.
Since the matrix is also antisymmetric, such that all eigenvalues are degenerate, there should be at least two zero eigenvalues.
These two zero modes correspond to the eigenvectors $(1,0,0,\sqrt{m+1},0,\ldots)$ and $(0,1,-\sqrt{m+1},0,0,\ldots)$ of the matrix in \cref{eq:H3}, which therefore correspond to the normalized eigenstates
\begin{equation}\label{eq:recurrence}
\widetilde\gamma_1''=\frac1{\sqrt{m+2}}\left( \widetilde\gamma_1+\sqrt{m+1}\,\widetilde\gamma_2' \right)
,
\qquad
\widetilde\gamma_2''=\frac1{\sqrt{m+2}}\left( \widetilde\gamma_2-\sqrt{m+1}\,\widetilde\gamma_1' \right),
\end{equation}
By using \cref{eq:modesH2,eq:modesinduction2} one gets
\begin{subequations}
\begin{align}
\widetilde\gamma_1''=& \frac1{\sqrt{2m+4}}
\left[
\gamma_{L1} + \gamma_{R2}
+
\sum_{n=1}^{m+1}
\kappa_{n-1}
\left(\gamma_{R,3n+1} - \gamma_{L,3n+2}\right)
+
\kappa_{n}
\left(\gamma_{L,3n+1} + \gamma_{R,3n+2}\right)
\right]
,
\\
\widetilde\gamma_2''=& \frac1{\sqrt{2m+4}}
\left[
\gamma_{R1} - \gamma_{L2}
-
\sum_{n=1}^{m+1}
\kappa_{n-1}
\left(\gamma_{L,3n+1} + \gamma_{R,3n+2}\right)
+
\kappa_{n-2}
\left(\gamma_{R,3n+1} - \gamma_{L,3n+2}\right)
\right]
,
\end{align}
\end{subequations}
using the fact that $\kappa_{n-1}=-\kappa_{n+1}$ and $\kappa_{n-2}=-\kappa_n$ in the second equation above, and reordering the terms one gets
\begin{subequations}
\begin{align}
\widetilde\gamma_1''=& \frac1{\sqrt{2(m+1)+2}}
\left[
\sum_{n=0}^{m+1}
\kappa_{n}
\left(\gamma_{L,3n+1} + \gamma_{R,3n+2}\right)
+
\kappa_{n-1}
\left(\gamma_{R,3n+1} - \gamma_{L,3n+2}\right)
\right]
,
\\
\widetilde\gamma_2''=& \frac1{\sqrt{2(m+1)+2}}
\left[
\sum_{n=0}^{m+1}
\kappa_{n}
\left(\gamma_{R,3n+1} - \gamma_{L,3n+2}\right)
+
\kappa_{n+1}
\left(\gamma_{L,3n+1} + \gamma_{R,3n+2}\right)
\right]
,
\end{align}
\end{subequations}
We thus completed the proof by induction, demonstrating that $H_{3(m+1)+2}$ has at least two energy zero modes in the form of \cref{eq:modesinduction}.
Again, due to the particle-hole symmetry, the number of positive energy levels $E>0$ is equal to the number of negative energy levels $E<0$.
Since the total number of eigenvalues $2N$ is even, the number of zero energy modes must also be even.
Moreover, direct numerical calculations indicate that there are only two zero energy modes in this case. 
Hence, we conclude that the Hamiltonian $H_{N}$ with odd $N=3m+2\ge2$ has two and only two zero eigenvalues.

The case $w=w'$ is realized when the distance between the Majorana modes at the left and right edges of the system is comparable with the distance between contiguous Majorana modes on the same side.

%\bibliographystyle{prsty_no_etal_titles_doi_preprint_noemph}
%\bibliography{bib}